\documentclass[a4paper,fleqn]{cas-dc}

\usepackage[numbers]{natbib}

\usepackage[mathscr]{eucal}
\usepackage{amsmath,mathtools}
\usepackage{amssymb}
\usepackage{lipsum}
\usepackage{algorithmicx}
\usepackage[ruled]{algorithm}
\usepackage[noend]{algpseudocode}
\usepackage{subfig}

\usepackage[switch]{lineno} 
\modulolinenumbers[2]

\newcommand*{\img}[1]{%
    \raisebox{0\baselineskip}{%
        \includegraphics[
        height=0.6\baselineskip,
        width=0.6\baselineskip,
        keepaspectratio,
        ]{#1}%
    }%
}

\def\tsc#1{\csdef{#1}{\textsc{\lowercase{#1}}\xspace}}
\tsc{WGM}
\tsc{QE}
\tsc{EP}
\tsc{PMS}
\tsc{BEC}
\tsc{DE}

\begin{document}
\let\WriteBookmarks\relax
\def\floatpagepagefraction{1}
\def\textpagefraction{.001}
\shorttitle{Contrastive analysis for scatterplot-based representations of dimensionality reduction}
\shortauthors{Marcílio-Jr et al.}

\title [mode = title]{Contrastive analysis for scatterplot-based representations of dimensionality reduction}                      



\author[1]{Wilson E. Marcílio-Jr}
\author[1]{Danilo M. Eler}
\author[1]{Rogério E. Garcia}


\address[1]{Faculty of Sciences and Technology, São Paulo State University (UNESP), Presidente Prudente, SP 19060-900, Brazil}





\cortext[cor1]{E-mails: wilson.marcilio@unesp.br, danilo.eler@unesp.br, rogerio.garcia@unesp.br} 




\begin{abstract}
Cluster interpretation after dimensionality reduction (DR) is a ubiquitous part of exploring multidimensional datasets. DR results are frequently represented by scatterplots, where spatial proximity encodes similarity among data samples. In the literature, techniques support the understanding of scatterplots' organization by visualizing the importance of the features for cluster definition with layout enrichment strategies. However, current approaches usually focus on global information, hampering the analysis whenever the focus is to understand the differences among clusters. Thus, this paper introduces a methodology to visually explore DR results and interpret clusters' formation based on contrastive analysis. We also introduce a bipartite graph to visually interpret and explore the relationship between the statistical variables employed to understand how the data features influence cluster formation. Our approach is demonstrated through case studies, in which we explore two document collections related to news articles and tweets about COVID-19 symptoms. Finally, we evaluate our approach through quantitative results to demonstrate its robustness to support multidimensional analysis.
\end{abstract}




\maketitle


\section{Introduction}



The analysis of high-dimensional datasets through dimensionality reduction (DR)~\citep{Paulovich_2008, Maaten_2008, McInnes2018} presents unprecedented opportunities to understand various phenomena. Using scatterplot representations of DR results, analysts inspect clusters to understand data nuances and features’ contribution to the layout organization in the projected space.

One promising strategy to analyze DR results, called contrastive analysis~\cite{Fujiwara2019, Le2019}, is understanding how clusters differ. Thus, the motivation is to find and comprehend the unique characteristics of each cluster. For instance, a tool for labeling textual data would benefit from contrastive analysis to highlight the differences among the clusters (e.g., topics)---these clusters would represent candidates for new classes because each cluster has its unique characteristics. Another important application is understanding which features describe two separated groups of patients after a medical experiment~\cite{Abid2018}.

Only a few works are focusing on providing contrastive analysis~\cite{Fujiwara2019, Le2019}. Instead, the literature presents a handful of studies that support the analysis of DR results through the visualization of global information~\cite{Turkay2012, Joia2015, Coimbra2016, Stahnke2016, Pagliosa2016}, emphasizing the importance of data features (attributes) given by DR techniques to organize the embedded space. A few main approaches~\cite{Turkay2012, Joia2015} use the principal components (PCs) from PCA~\cite{Jolliffe1986} to find which features contribute to cluster formation. However, these contributions do not emphasize the unique characteristics of the clusters. On the other hand, ccPCA~\cite{Fujiwara2019} finds contrastive information (i.e., the unique characteristics) for each cluster through the systematic application of contrastive PCA~\cite{Abdi2007}. ccPCA's main limitation comes from PCA, i.e., the prohibitively run-time execution for high-dimensional datasets---it takes $O(m^3)$ on the number of dimensions.
Another approach, ContraVis~\citep{Le2019}, is only applied to textual data and cannot help interpret DR layouts since it is already a dimensionality reduction approach. Furthermore, the study does not present a strategy to understand the terms' contribution to the visual space layout organization.

In this work, we propose cExpression, an approach to analyze DR results using contrastive analysis along with a carefully designed visualization technique. More precisely, we use statistical variables (p-values and $t$-scores) to find the most distinctive features of clusters (t-scores) and the confidence of the results (p-values). Although t-test is a common method for feature selection in machine learning and in bioinformatics for defining cell types, it is not well-explored to analyze dimensionality reduction results. In our visualization design, users can interact with scatterplot representations of multidimensional datasets to visualize the clusters' summaries---designed after the definition of several requirements. We use focus+context interaction on a bipartite graph to communicate the relationship between t-scores and p-values. The focus+context interaction helps users explore a higher amount of information while inspecting small-multiples of features' histograms. A heatmap representation of the most distinctive features for each cluster also helps to overview the structures. Finally, we propose an encoding strategy to simultaneously communicate the distribution of feature values in the scatterplot representation. 

As we demonstrate in the numerical experiments, cExpression can be applied to various data types and is scalable to handle big datasets with thousands of dimensions. While our approach's computational components help generate contrastive information rapidly, our visualization design is simple and effective to analyze even complex textual data.

In summary, our contributions are:

\begin{itemize}
\item A strategy to analyze and interpret dimensionality reduction through clusters using contrastive analysis;
\item Novel visualization strategies to analyze the relationship among statistical variables and simultaneously visualize various features in the scatterplot;
\item An annotated dataset of COVID-19 tweets retrieved from March 2020 to August 2020.
\end{itemize}	

This work is organized as follows: Section~\ref{sec:related-works} presents the related works; Section~\ref{sec:cexpression} delineates our methodology accompanied with motivation and the visualization design; Section~\ref{sec:case-studies} shows the case studies; Section~\ref{sec:evaluation} shows numerical evaluation; Section~\ref{sec:discussions} presents discussions about the work; the work is concluded in Section~\ref{sec:conclusion}.

\section{Related Works}
\label{sec:related-works}

To support analysis of dimensionality reduction (DR) results, layout enrichment strategies~\cite{NonatoAupetit} unite visualization approaches and valuable information extracted from the data on high-dimensional space concerning low-dimensional representations, usually on $\mathbb{R}^2$. Examples include using bar charts and color encoding to understand three-dimensional projections~\citep{Coimbra2016} or encoding attribute variation using Delaunay triangulation to assess neighborhood relations in projections on $\mathbb{R}^2$~\citep{Silva2015}. Another interesting work, proposed by Tian et al.~\cite{Tian2021}, extends Silva et al.'s~\cite{Silva2015} by aggregating the explanatory mechanisms using a visual analytics system, together with a new local explanation method (variance ratio) that describes the dimensionality of local neighborhoods. Probing Projections~\citep{Stahnke2016}, for example, depicts error information by displaying a halo around each dot in a DR layout besides providing interaction mechanisms to understand distortions in the projection process. The majority of the works use traditional statistical charts to visualize attribute variability~\citep{Pagliosa2016}, neighborhood and class errors~\citep{MarcilioJr2017}, or quality metrics~\citep{Kwon2018}. For instance, Martins et al.~\citep{MarcilioJr2017} use space-filling techniques to help users reason about the influence on neighborhood preservation and other quality aspects of parameterized projections.

More related to our work are techniques that find important features given clusters of data points. For example, the Linear Discriminative Coordinates~\citep{Wang2017} technique uses LDA~\citep{Izenman2008} to produce cohesive clusters by discarding the least important features. Joia et al.~\cite{Joia2015} use PCA to find the most important features through a simple matrix transposition to later visualize feature names as word clouds within each cluster region---such an approach tends to be influenced by classes with a higher variation. Our approach, however, surpasses these problems by comparing the differences between distributions and returning the confidence levels associated with such differences. Another work, proposed by Turkay et al.~\cite{Turkay2012}, also uses the principal components computed by PCA to obtain MDS's representative features~\citep{Kruskal1978}. The problem with using PCA is that it focuses on the global characteristics of datasets, making it challenging to analyze the unique characteristics of clusters as we address them using contrastive analysis.

Recently, Fujiwara et al.~\cite{Fujiwara2019} proposed the contrastive cluster PCA (ccPCA) technique that finds the most important features for each cluster in a projection. Fujiwara et al.'s approach is different from Joia et al.'s and Turkay et al.'s works to provide a way to understand which features highly contribute to the differentiation of clusters. ccPCA's main limitation is the prohibitively run-time execution for datasets with many dimensions, such as document collections~\cite{Lopes2007, Wang2013, Eler2018, Felix2018}, gene expression~\citep{Murie2008, Robinson2009, Ritchie2015, Hollt2018}, or filter activations~\cite{MarcilioJr2020}. Finally, there is no consistent way to visually relate the distribution of values with the feature's contribution returned based on contrastive PCA. Unlike ccPCA, our proposal successfully analyzes datasets with very high dimensionality while presenting more consistent results---as shown in the following section. Further that, our approach yields more interpretable results since it is based upon well-known statistics measures. Another exciting work, proposed by Le et al.~\citep{Le2019}, combines topic extraction to visualize the unique features from document collections. Although their technique can differentiate clusters well, it cannot be used to analyze other dimensionality reduction approaches, and it is limited to textual data. Our approach is uncoupled from specific dimensionality reduction techniques and can be used to analyze various data types.

In the following section, we detailed cExpression, which comprises a method for retrieving contrastive information and an associated visualization tool.

\section{cExpression---Tool for contrastive analysis of DR results}
\label{sec:cexpression}

\begin{figure}[h!]
\centering
\includegraphics[width=\linewidth]{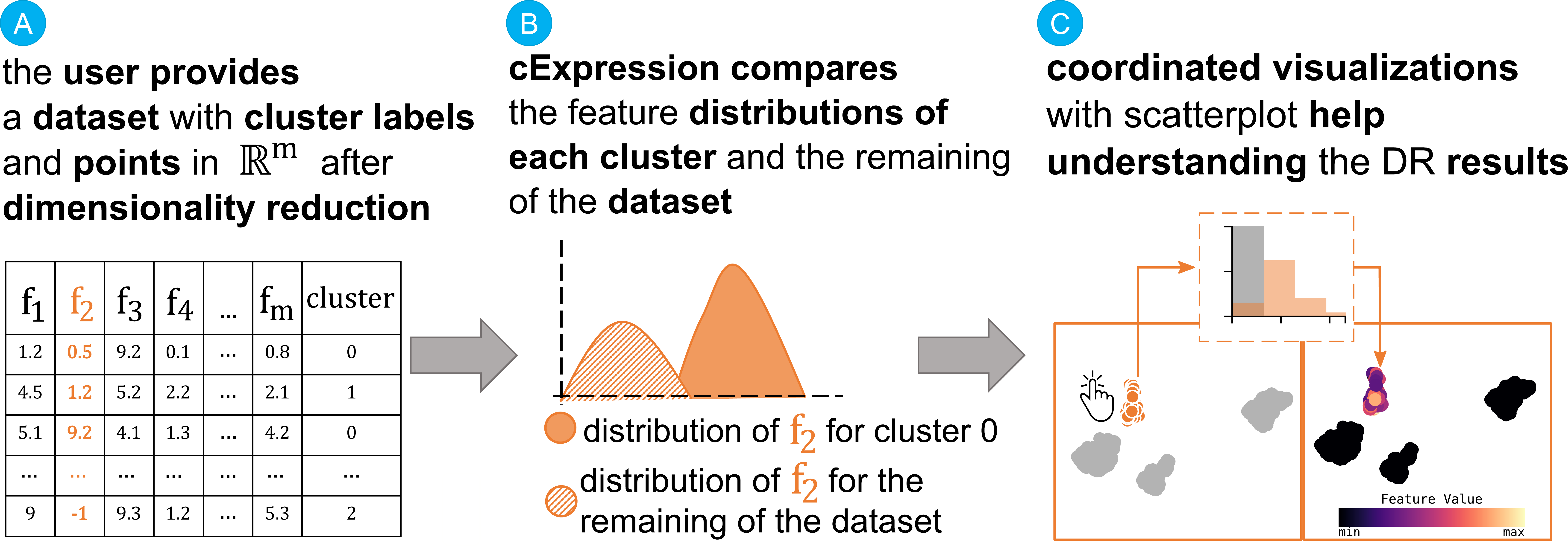}
\caption{ The cExpression pipeline. The user provides a high-dimensional dataset with $\mathbb{R}^2$ coordinates after dimensionality reduction and cluster labels (A). The clustering labels and high-dimensional data are used to compute the contrastive information for each par of cluster and feature (B). The data generated from this process is used to create visualization approaches that highlight the unique characteristics of each cluster (C).}\label{fig:pipeline}
\end{figure}

Before detailing our approach in the following sections, Fig.~\ref{fig:pipeline} shows the workflow for using cExpression to interpret clusters after dimensionality reduction. First, the user has to preprocess a high-dimensional dataset by applying a dimensionality reduction technique and annotating the clusters perceived in the visual space, which results in the state (A). Then, our approach uses t-test to compute a measure of deviation (t-score) and associated confidence level (p-value) for each pair of feature and cluster---for instance, the state (B) in Fig.~\ref{fig:pipeline} represents two distributions associated to the feature $f_2$, one for the cluster $0$ and another for the remaining of the dataset. The comparison using t-test answers the question of ``Is this feature a unique characteristic for cluster 0?''. Finally, the data generated from this process feeds a visualization tool (C) that uses coordinated views to help in the exploratory process.

\subsection{Computing constrastive information}
\label{sec:methodology-motivation}

Our method is based on comparing distributions of values of different clusters and using $t$-test to verify if the differences among these distributions are enough to describe clusters. Given a dataset $X$, let $X_f^c$ and $X_f^{c'}$ be the values of feature $f$ for the data observations in the cluster $c$ and the values of the features $f$ for the data observations \textit{not} in cluster $c$ $(c')$---$c$ and $c'$ are clusters defined in the visual space from which users want to find unique characteristics. Then, we compute the summary statistics to get the $t$ value for the $t$-test: average and variance for $X_f^c$ and $Y_f^{c'}$, defined as $\overline{X_f^c}$, $\overline{X_f^{c'}}$ and $Var(X_f^c)$, $Var(X_f^{c'})$. Finally, the $t$-statistic can be calculated as it follows:

\begin{equation}
\label{eq:t-statistic}
t = \frac{\overline{X_f^c} - \overline{X_f^{c'}}}{\sqrt{ \frac{Var(X_f^c)}{|X_f^c|} + \frac{Var(X_f^{c'})}{|X_f^{c'}|} }}
\end{equation}

\begin{figure}[h!]
\centering
\includegraphics[width=0.8\linewidth]{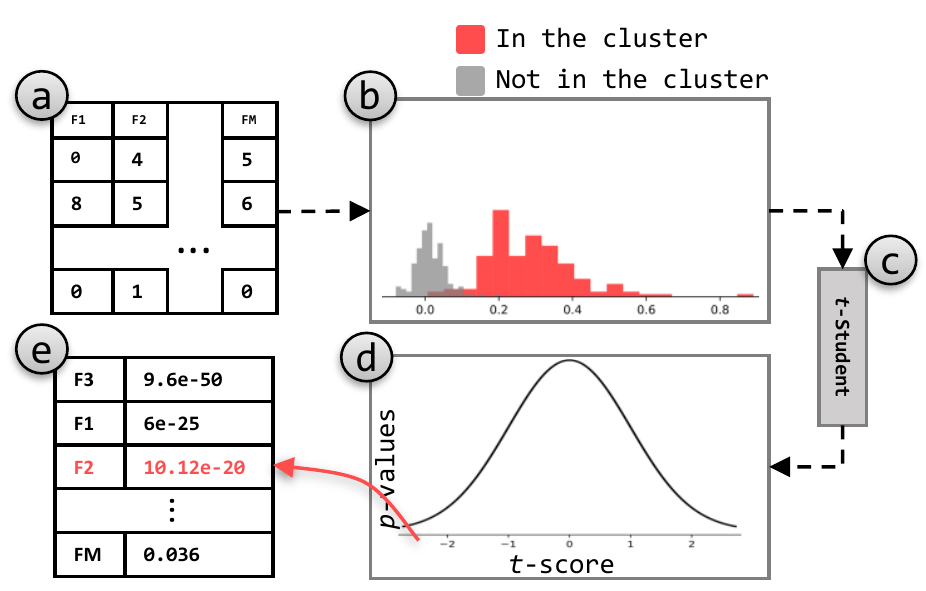}
\caption{Defining distinctive features. The feature's distribution (\textbf{a}) in a cluster of interest is compared to its distribution on the remaining dataset (\textbf{b}). Then, the $t$-Student test (\textbf{c}) returns the probability of these two distributions to be different (\textbf{d}). Finally, the features are arranged in decreasing order according to their $t$-score to communicate discriminative power (\textbf{e}).}\label{fig:methodology}
\end{figure}

\noindent in which $|\bullet|$ stands for the size of the sets. Having the $t$-statistic, we use a statistical library to find the $p$-value, which corresponds to the probability of rejecting a null hypothesis. In other words, the $p$-value represents the probability in which we can assume that the distribution of values in cluster $c$ for feature $f$ ($X_f^c$) and the distribution of values not in cluster $c$ for feature $f$ ($X_f^{c'}$) are equal. Notice that, fixing a cluster $c$ and a feature $f$, smaller $p$-values stands for high importance since it means (by the null hypothesis) that the distributions are different---the t-scores are measures of standard deviation. In this case, very high or very low (negative) t-scores are associated with very small $p$-values and are found in the tails of the $t$-distribution---as illustrated in Fig.~\ref{fig:methodology}(\textbf{d}). A similar approach to compute explanations, the variance-based explanation~\cite{Silva2015, Tian2021} computes the relative variance of a neighborhood the entire dataset, thus, emphasizing how dimensions contribute to similarity within local neighborhoods---such an explanation approach emphasizes which dimension makes their neighbors similar while the $t$-score is a measure to understand how much two distributions deviate from each other. The main difference between these two approaches consists of the way they find these features' relevance. While the $t$-score compares disjoint sets, the variance-based explanation looks at the contribution to neighborhoods over the entire dataset.

To interpret clusters using contrastive analysis, we perform the process described above for each pair of cluster and feature of the dataset. Then, t-scores are used as a ranking metric. It is important to emphasize that these clusters must be perceived in the visual space, so the insight generated from the exploratory process is consistent with the organization visualized in the scatterplot. Another important aspect is that the cluster labels---encoded by different colors and responsible for defining the clusters---are not computed by our tool. Instead, users have to provide their own processed file with cluster labels and ($x,y$) coordinates resulted from dimensionality reduction.

Fig.~\ref{fig:methodology} illustrates the whole idea for a fixed cluster and a fixed feature. From the dataset in (\textbf{a}), the distributions of values for the cluster and feature are generated (\textbf{b})---red encodes the feature distribution of data samples in the cluster, and gray encodes the distribution of data samples outside the cluster. Based on these distributions, we used the $t$-Student test (\textbf{c}) to ask for the probability ($p$-value, \textbf{d}) in which the distributions are similar. Finally, ordering the cluster's features in an decreasing way of absolute $t$-score will lead to the most defining features of the cluster being placed on top.

\begin{figure}[htb!]
\centering
\includegraphics[width=\linewidth]{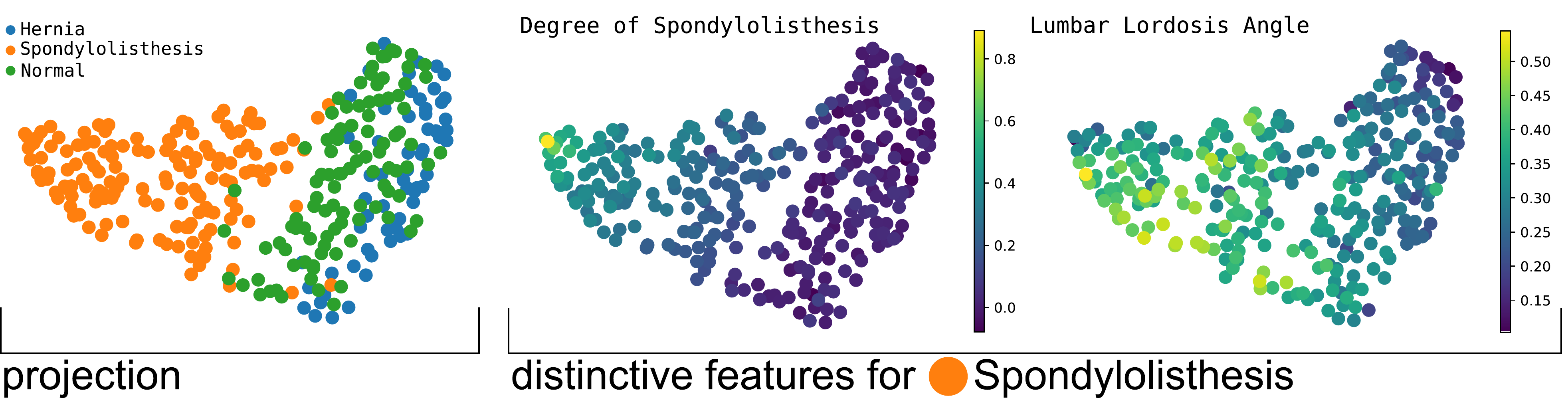}
\caption{UMAP projection of the \textit{Vertebral Column} dataset and two distinctive features for the Spondylolisthesis class.}\label{fig:vertebral}
\end{figure}

Let us apply the concept delineated above on a multivariate dataset. Fig.~\ref{fig:vertebral} (left) shows a UMAP~\citep{McInnes2018} projection of the Vertebral dataset, which consists of $310$ data instances described by six bio-mechanical features derived from the shape and orientation of the pelvis and lumbar spine: \texttt{pelvic incidence}, \texttt{pelvic tilt}, \texttt{lumbar lordosis angle}, \texttt{sacral slope}, \texttt{pelvic radius}, \texttt{degree of spondylolisthesis}---see \textbf{Supplementary File} with a case study with this dataset. The class of interest corresponds to patients with Spondylolisthesis, a disturbance of the spine where a bone (vertebra) slides forward over the bone below it.

One critical feature for patients with Spondylolisthesis is the  \texttt{degree of spondylolisthesis}, which is known to be high in those patients. It is reasonable to think that an algorithm that tries to find important features would select the  \texttt{degree of spondylolisthesis} for defining such a class. To investigate \textit{cExpression} on this task, we explore the distribution of values of the important features throughout the projection. Fig.~\ref{fig:vertebral} (right) shows the two most important retrieved by cExpression. The scatterplot showing the normalized feature values also confirms that \texttt{degree of spondylolisthesis} is a reasonable candidate for a distinctive feature. That is, data instances from class \img{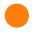} present much higher values for such a feature than data instances from classes \img{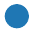} and \img{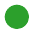}.


\subsection{Visualization Design}
\label{sec:visualization-design}

\begin{figure*}[h!]
\centering
  \frame{\includegraphics[width=\linewidth]{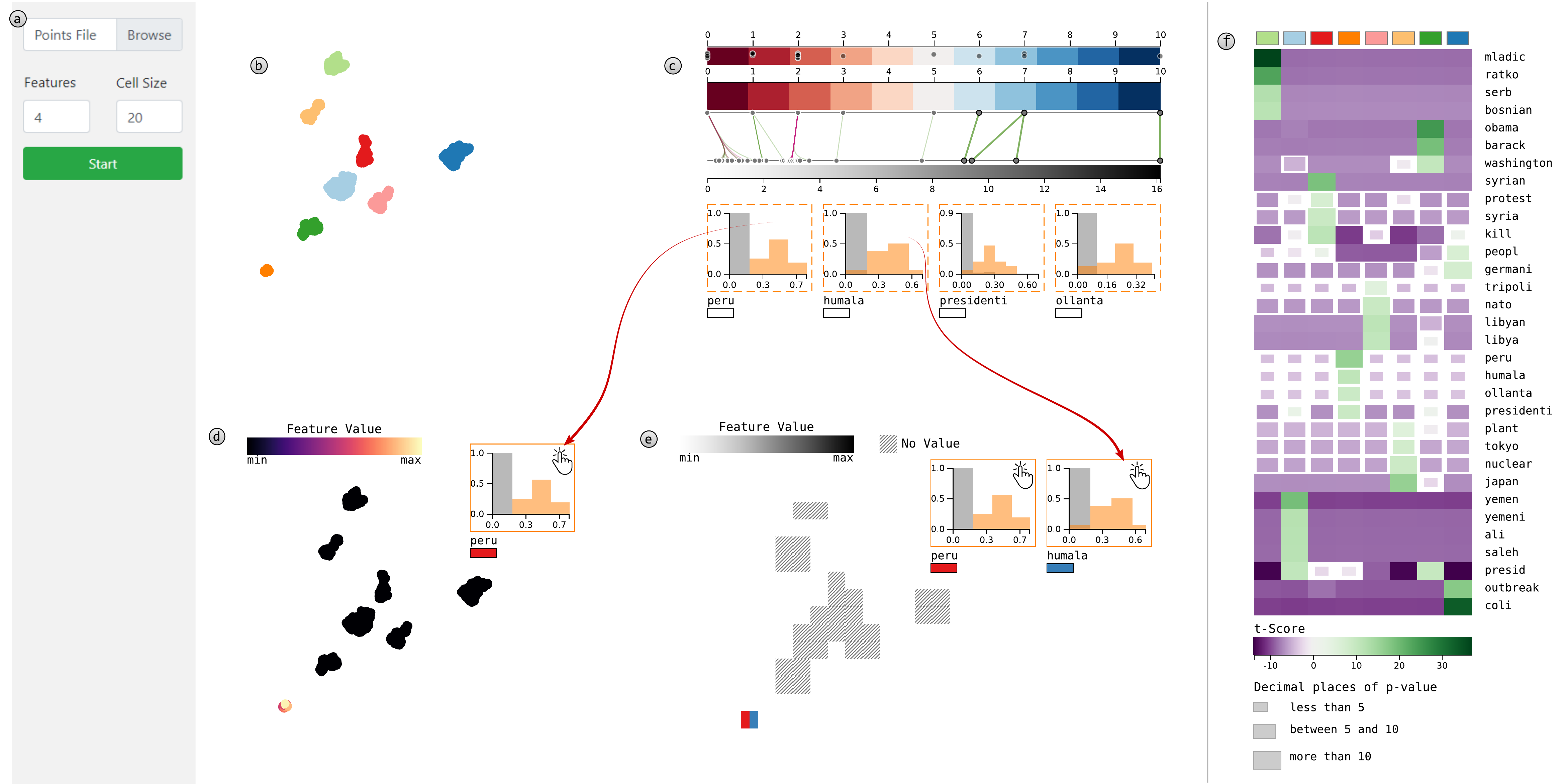}}
  \caption{The cExpression visualization tool. Users provide high-dimensional points with annotated clusters and positions onto $\mathbb{R}^2$ (a). The $\mathbb{R}^2$ coordinates are used for the scatterplot layout (c), from which the analysis starts. Users click on points of a cluster to investigate their contrastive information. For instance, the bipartite graph (c) shows the distinctive terms for the dark-orange cluster. There is also an interaction mechanism to visualize the feature values on the scatterplot visualization (d, e). Finally, an overview analysis can be achieved through the heatmap encoding.}
	\label{fig:tool}
\end{figure*}

We designed a few visualization strategies to analyze dimensionality reduction results with \textit{cExpression}. The tool in Fig.~\ref{fig:tool} shows two main views: the scatterplot view and the heatmap view. Firstly, users provide the high-dimensional dataset and the points after dimensionality reduction ($\mathbb{R}^2$), define the cell size for the cell-based encoding visualization, and specify the number of features automatically inspected using the summary visualization (\textbf{a}). The scatterplot representation of the projected dataset (\textbf{b}) serves as the basis for the interaction with the bipartite graph (\textbf{c}), representing the relationship among the statistical variables. In Fig.~\ref{fig:tool}, for example, the bipartite graph shows information for the cluster \img{figs/motivation/orange}. Users can toggle the distribution plots (\textbf{d} and \textbf{e}) to inspect the scatterplot representation distribution. The heatmap view (\textbf{f}) summarizes the feature importance, where a diverging color scale from purple to green encodes the t-scores. The sizes of the heatmap' tiles encode the number of decimal places of the associated p-value. High t-scores (represented by high color saturation) are associated with lower p-values (greater tile sizes) in the heatmap visualization.
We explain the decisions about the visualization encodings by first delineating the task and visual requirements in the next section.

\subsubsection{Visual and task requirements}

We have followed the layout enrichment~\citep{NonatoAupetit} strategy of scatterplot representations after dimensionality reduction to design our visualizations. Thus, the following task requirements (TRs) help us to understand the cluster formation in DR results (projection):

\begin{itemize}
\item\textbf{TR1:} Compare a cluster against the remaining of the dataset;
\item\textbf{TR2:} Compare two selected clusters.
\end{itemize}

The task requirements (TRs) help users to understand the projection structures by comparing how clusters differentiate. To achieve these tasks, we delineate the minimum requirements that our visualization must comprise:

\begin{itemize}
\item\textbf{VR1:} Visualize the importance of a feature;
\item\textbf{VR2:} Visualize the distribution of values of the comparing components;
\item\textbf{VR3:} Know the feature name;
\item\textbf{VR4:} Assess how trustful the result is;
\item\textbf{VR5:} Understand the organization of features distribution in the DR layout.
\end{itemize}

These visualization requirements (VRs) are analyzed together during exploratory data analysis. That is, knowing the most discriminative features for a cluster (VR1 and VR3), users understand intra-cluster differences from the remaining of the dataset (VR2) and how their values are distributed throughout the DR layout (VR5). VR4, related to the p-values, helps users on the statistical relevance of the difference.

Lastly, since tasks TR1 and TR2 are not new when dealing with scatterplot visualizations~\cite{Sedlmair2013, Micallef2017, Sarikaya2018b}, other works address the same problem. One similar work~\cite{Rauber2015} uses feature selection to re-project high-dimensional data onto $\mathbb{R}^2$ based on feature selection to induce cluster separation and help in classification tasks. Instead, we focus on contrastive analysis to explain which features are important for cluster formation---we are interested in improving the understanding but not improving the projection itself.

\subsubsection{Visualizing the feature importance}
\label{sec:feature-visualization}

\begin{figure}[h!]
\centering
\includegraphics[width=\linewidth]{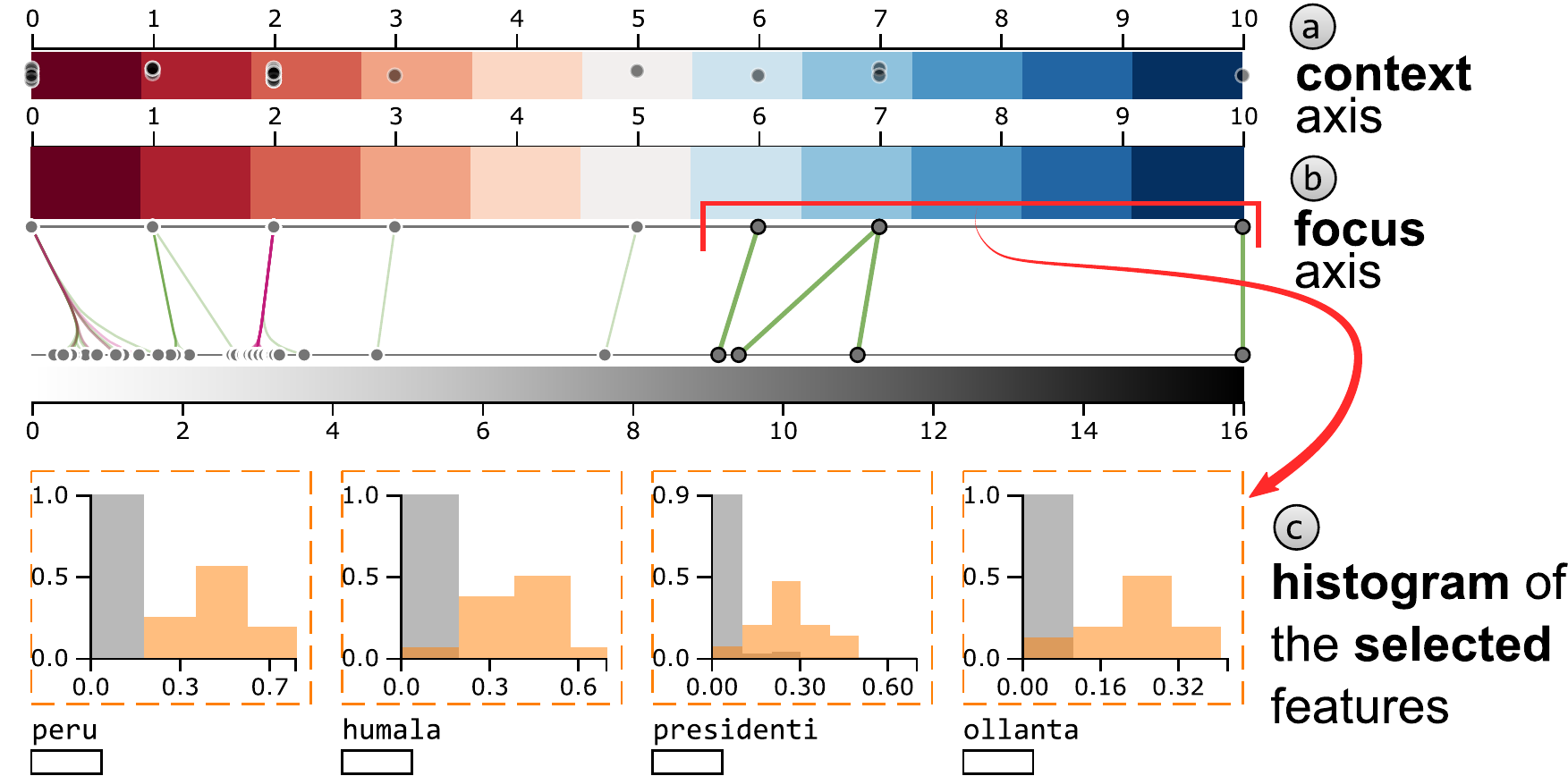}
\caption{The bipartite graph encoding feature importance. A red-to-blue and continuous color scales encode the p-values and t-values, respectively. Notice that, we visualize the number of decimal places after zero instead of visualizing the p-value itself. For focus+context interaction, the first axis (a) gives the context of the available features while the second axis (b) correspond to the focus---in this case, the are the same. The links between the focus axis and the t-score represent the relationship between p-values and t-scores, in which green colors represent positive values and red colors represent negative values. Finally, the four most distinctive features are pre-selected and visualized as a histogram plot (c).}\label{fig:summary-importance}
\end{figure}

To compactly represent the VR1-4 requirements, we visualize the importance of the features as the position over a horizontal axis while providing the distribution plots for each visualized feature. Fig.~\ref{fig:summary-importance} shows the visualization design, where the color hue for the distribution plots corresponds to cluster \img{figs/motivation/orange} and the operation illustrated in Fig.~\ref{fig:tool} (c).

The features' importance is encoded by a line segment, representing the relationship between p-values and t-scores. The axes (\textbf{a}) and (\textbf{b}) in Fig.~\ref{fig:summary-importance} present the same information, i.e., the number of decimal places after zero of each p-value. These two axes are central to our focus+context interaction, detailed in Section~\ref{sec:interaction}. The gray circles in the context axis (\textbf{a}) represent the analyzed features whose horizontal position encodes the p-values---since there is no information to encode on vertical position, we assign a random value to reduce overlapping among circles. Notice that with this design, we accomplish requirements VR1 and VR4. However, we still need to comprise other visualization requirements to differentiate between the features while understanding the confidence in the feature differentiation. Finally, the bipartite graph visualization in Fig.~\ref{fig:summary-importance} shows the $min(m, 50)$ most important features to reduce overplotting, where $m$ stands for the dimensionality of the dataset. We focus on a selected number of features independently from the dataset dimensionality since we empirically found that the most helpful information remains on the few top ones. Further that, the use cases show how this approach is adequate for even textual data, which usually contains hundreds of dimensions.

A red-to-blue color scale help users to identify the feature confidence---blueish colors encode more confidence (or lower p-values). Besides the confidence in the difference represented by the p-values, we encode the difference between distributions (t-scores) using another axis. The t-score represents the deviation between the distributions of a feature for the analyzed cluster and the remaining of the dataset. The relationship among these statistical variables is shown through line segments, and the color indicates the signal of t-score (pink for negative and green for positive). These line segments are drawn with higher transparency when they are not selected, and the edge bundling algorithm is employed to reduce the over-plotting---see the effect on the left part of Fig.~\ref{fig:summary-importance}. As we will show in the following section, the rectangles below the features' names help dynamically assign color hues to features when exploring their distribution on the scatterplot representation.

The requirements VR2 and VR3 are fulfilled with a distribution plot enriched with other information. The distribution of the values is encoded using a histogram (see  Fig.~\ref{fig:summary-importance}(\textbf{c})). As in Fujiwara et al.'s work~\citep{Fujiwara2019}, the $y$ axis encodes the relative frequency of the bins, that is, the number of data observations with a particular value divided by the number of data observations in the cluster (for histograms of the cluster) or outside the cluster (for histograms of the remaining data). The color bars of the distribution plots (and their borders) assume the corresponding cluster's color. In contrast, the distribution plots of the values for the remaining dataset receive gray. For example, the distribution plots (\textbf{c}) have the same color (\img{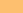}) as the selected cluster (\textbf{b}) in Fig.~\ref{fig:tool}.

\subsubsection{Visualizing feature value distribution}
\label{sec:visualizing-similarity}

\begin{figure}[h!]
\centering
\includegraphics[width=\linewidth]{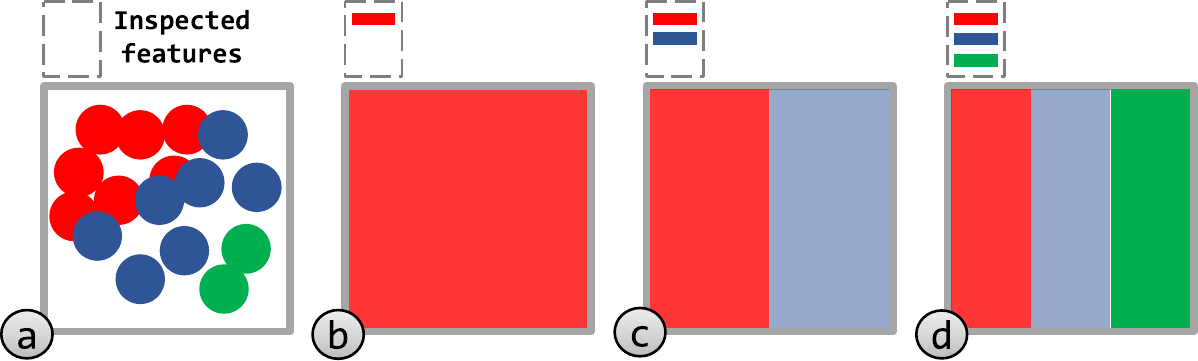}
\caption{Steps for construction of the cell-based encoding. The sum of each feature is retrieved for each cell (a). Then, according to user demand, the cell is divided to communicate the features. The space dedicated for a feature corresponds to the number of features present in the cell. For instance, one feature in (b), two features in (c), and three features in (d).}\label{fig:cell-values}
\end{figure}

Visualizing the distribution of feature values helps to understand the influence of features in the DR layout. That is, in Figs.~\ref{fig:tool} (d) and (e), the user interacts with the tool to visualize the feature values of features \texttt{peru} and \texttt{humala}. A continuous color map solves the problem for a single feature (e.g., see Fig.~\ref{fig:tool} (d)). However, a higher number of features requires further attention to maintain the similarity relations and the analysis power of proximity-based scatterplot representations (e.g., see Fig.~\ref{fig:tool} (e)). So, inspired by Sarikayra et al.'s~\cite{Sarikaya2018} work to visualize classification metrics of protein surfaces, we discretize the projected space in cells of fixed size (in pixels) to plot the contribution of each feature to the cell. Fig.~\ref{fig:cell-values} exemplifies the construction of the encoding for three features and only one cell.

Fig.~\ref{fig:cell-values}(\textbf{a}) shows a cell in the projected space that contains a hypothetical number of 16 data points with color encoding the most defining feature for each data point (\img{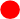}, \img{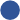}, \img{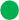}). A continuous color scale is a simple yet effective option to inspect only one feature (\textbf{b}) since the markers only change according to the hue. The problem comes when inspecting more than one feature while trying to maintain the organization imposed by the dimensionality reduction technique on the resulting layout. Thus, we use the grid structure to group similar data points---similar as in the SADIRE~\cite{MarcilioJr2020_SADIRE} sampling technique---and divide the space according to the number of features expressed inside each cell. In (\textbf{c}), after adding feature \img{figs/inline/feature2.png} to the inspection, the cell is divided and colored according to the features' color-encoding.
Moreover, color opacity is used to communicate feature expression inside the cell---the most opaque segment for a feature indicates the highest expression for such a feature. The same happens when the feature \img{figs/inline/feature3.png} is added to the analysis (\textbf{d}) when dividing the cell into three parts.

\begin{figure}[!htb]
\centering
\includegraphics[width=\linewidth]{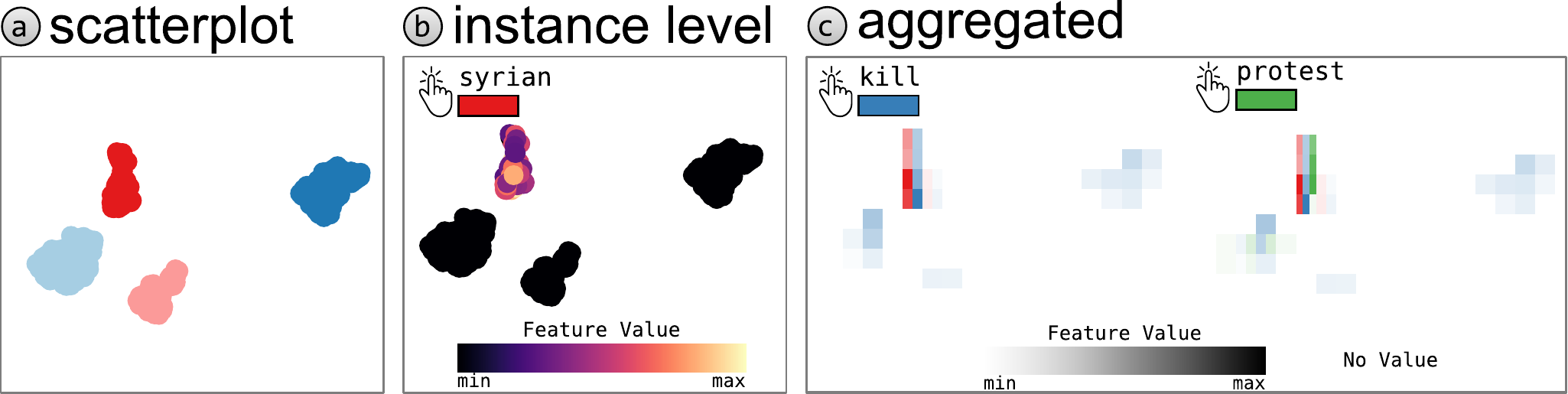}
\caption{Cell-based visualization for feature distribution on the scatterplot. After inspecting the distinctive features on the scatterplot (a), users can visualize the feature values using cell-based visualizations. For instance, when one feature is selected (b, \texttt{syrian}), a simple continuous color scale is sufficient. For more than one feature (c, \texttt{syrian, kill, protect}), cell-based encoding is employed.}
\label{fig:exemplification-scatter-distribution}
\end{figure}

Fig.~\ref{fig:exemplification-scatter-distribution} illustrates the encoding strategy discussed above. We selected four clusters (\textbf{a}). After inspecting cluster \img{figs/inline/feature1.png}, the selection of \texttt{syrian} shows where the term is mostly expressed (notice that the colors representing the cluster and the term itself are not related) using a color scale, which helps users to investigate where the documents are talking about \texttt{syrian}. The aggregated encoding uses color saturation to communicate the expression level in regions defined by cells of a fixed size (defined by users) to visualize the co-expression of terms. Notice that the space inside each cell is divided by the number of features being inspected.

The steps of Fig.~\ref{fig:exemplification-scatter-distribution} are achieved through user interaction with the distribution plots. As shown in Fig.~\ref{fig:tool}(\textbf{d}-\textbf{e}), borders with straight lines indicate the feature inspection, and users use combinations to visualize multiple values. Notice that cells in which no data observations express any feature would not be perceived in the visualization using the opacity strategy as discussed above. We used a texture (\img{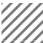}) to fill these cells and to maintain the general structure of the projected dataset. Lastly, the division of each cell corresponds to the number of expressed features in the cell---that is, if three features are selected but only two expressed, the cell is divided into two equal parts.

Another important thing to mention about the cell-based encoding is the trade-off between computational performance and quality of explanations related to the cell size in pixels. As we augment the cell size, finding which points belong to a cell becomes faster, although the resemblance to the original scatterplot decreases. The contrary is also true. As we reduce the cell size, the cell-based encoding approaches more the scatterplot structure. However, it adds a computational burden---a small cell size also prejudices the visualization of the color encoding. To tackle these problems, we use a KD-tree to identify which points belong to a cell efficiently and use a cell size of 20x20 pixels---we find it a good compromise between the visualization of feature distribution and scatterplot structures. Finally, users can define the cell size before generating the visualizations (see Fig.~\ref{fig:tool}).



\subsubsection{Comparing clusters}

Besides comparing a cluster of interest to the rest of the dataset, it is helpful to understand the differences between two clusters. For instance, to understand why subclusters are formed.

\begin{figure}[!htb]
\centering
\includegraphics[width=\linewidth]{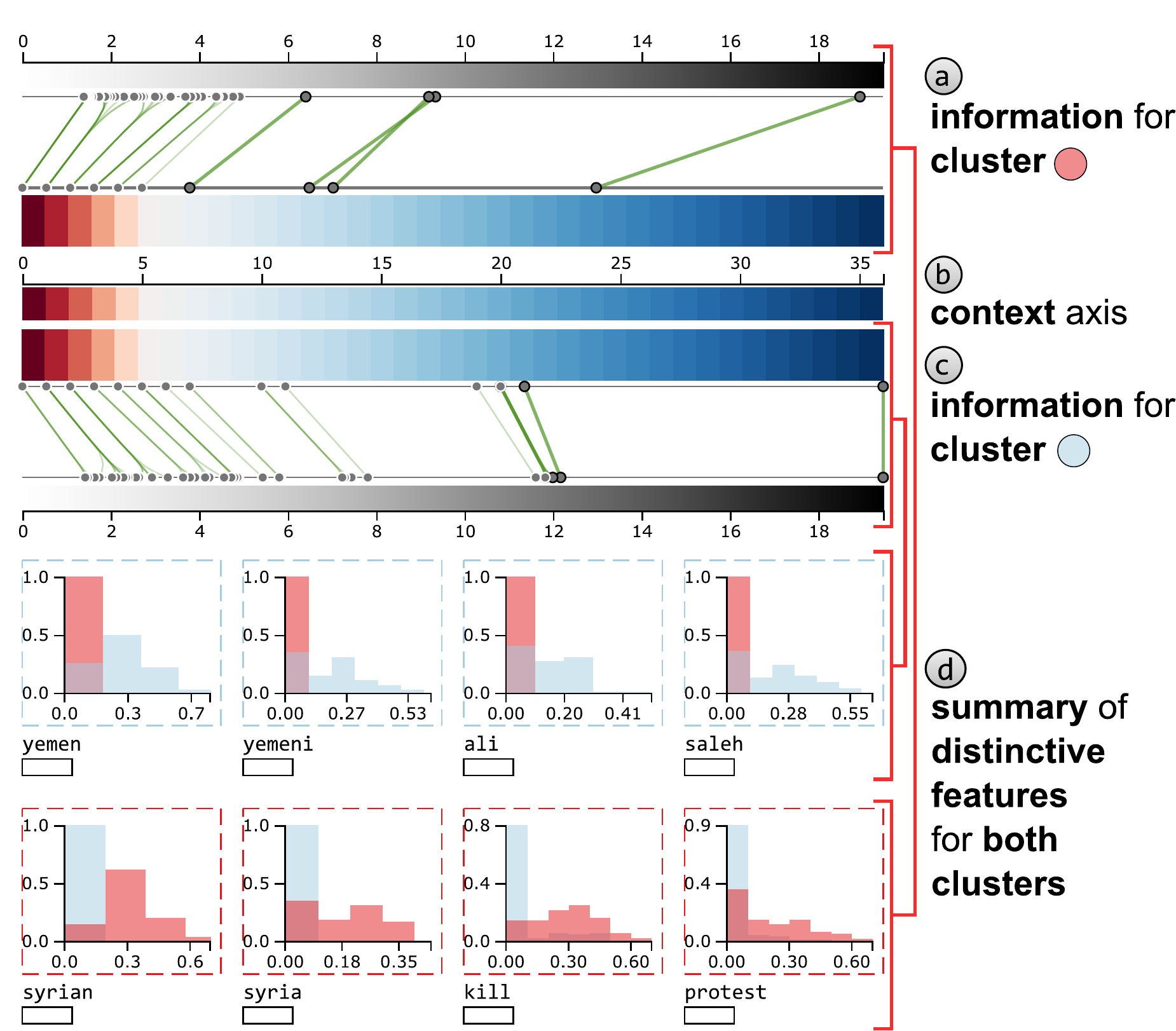}
\caption{Comparison between two clusters. To compare two clusters, the visualization strategy of Fig.~\ref{fig:summary-importance} is used to communicate the information of two clusters. Essentially, while the information for the first cluster (light-blue in the figure) is encoded as usual (c), the information for the second is positioned on top (a), and the context axis assumes the middle of the visualization (b). In addition, the four most distinctive features for both clusters are shown (d).}
\label{fig:contrasting-clusters}
\end{figure}

For comparison, users select two clusters of interest on the scatterplot view. In this case, the contrastive information is retrieved from two arbitrary clusters $c_a$ and $c_b$. Fig.~\ref{fig:contrasting-clusters} shows an augmented version of the visualization of Fig.~\ref{fig:summary-importance}, where the information about each cluster is mirrored---the color scale representing the p-values is constructed based on the clusters with the longer range. When comparing clusters, the information provided to contrast clusters to the remaining dataset is presented to both clusters in comparison (such as distribution plots, p-values, and t-scores axes).

\subsubsection{Summary View}

Although our technique emphasizes differences of the dataset even for a reasonable amount of features, it would be interesting to understand more about a cluster structure, visualizing the importance of more features for some datasets. A heatmap provides an overview of the feature importance and visualizes the statistics used by our technique, as shown in Fig.~\ref{fig:tool}(\textbf{f}). A purple to green color scale encodes the t-score, and the dimensions of the tiles encode the p-values. It is important to remember that t-scores are measures of standard deviation and p-values are probabilities. Both statistics are associated with standard t-distribution. This distribution relates standard deviations with probabilities and allows significance and confidence to be attached to t-scores and p-values. For this particular example, Fig.~\ref{fig:tool}(\textbf{f}) shows how the distinctive features help explain each cluster through contrastive analysis. That is, none of the features with green tiles are shared across columns.

\subsubsection{Interaction mechanisms}
\label{sec:interaction}

As shown in Fig.~\ref{fig:summary-importance}, the four most distinctive features are presented when inspecting a cluster. However, users might inspect other features based on their p-value. Such an interaction is carried out using a selection box, as illustrated in Fig.~\ref{fig:summary-importance} (smallest box). All of the corresponding features inside the box are detailed inspected by the visualization of their respective distribution plots. The line segments of the bipartite graph are also updated to communicate which features are selected.

\begin{figure}[!htb]
\centering
\includegraphics[width=\linewidth]{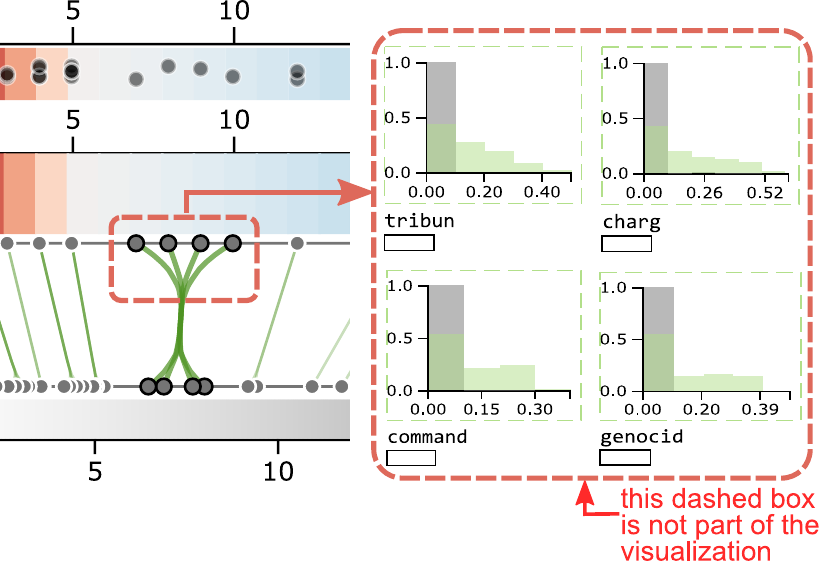}
\caption{Users can inspect the histograms of available features through lasso selection over the points encoding p-values.}
\label{fig:interaction-selection}
\end{figure}

Another differential of our tool is assigning colors to the features being inspected in the scatterplot. While users can freely visualize many distribution plots, there is a limit in the number of colors that humans can differentiate~\citep{Ward_2002}. With a limit of ten features to be simultaneously visualized, users can toggle the distribution plots. Fig.~\ref{fig:exemplification-scatter-distribution}(\textbf{c}) exemplifies this operation for the terms \texttt{kill} and \texttt{protest} and their result in the scatterplot representation using our encoding.
.



\begin{figure}[!htb]
\centering
\includegraphics[width=\linewidth]{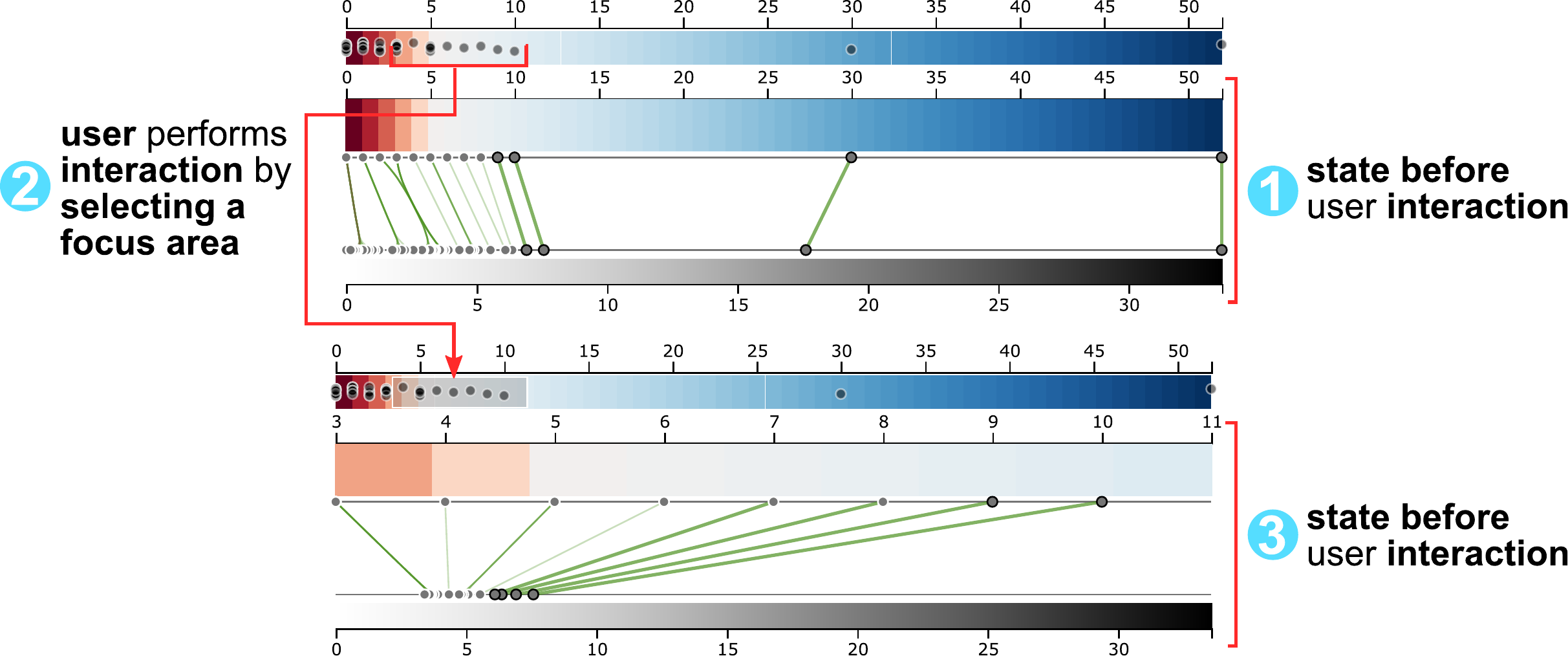}
\caption{Focus+context operation. The visualization starts with the context and focus axes as the same (1). Then, users can select a range of the context axis to focus analysis (2), resulting in an update of the focus axis (3).}
\label{fig:focus_context}
\end{figure}

The selection mechanism helps users to understand other aspects of the clusters by inspecting many features. However, when these features present a similar p-value, such an interaction is not enough to select interest features. In other words, the visual space dedicated to various features could be too small. We employ focus+context~\citep{Munzner2015} interaction on the p-value axis to decrease such an issue, as shown in Fig.~\ref{fig:focus_context}. On top of the axis showing the p-values, another axis corresponds to the context, i.e., all of the available information. At first, the focus axis corresponds to the context axis (1). Then, users specify a range where the focus axis will be defined (2), as illustrated in the figure by a red arrow. The focus selection will make the axis of p-values show only the information inside the selection box (3). Finally, such a change in the focus induces a change in the features visualized as distribution plots.

\subsection{Implementation}

The visualization metaphors were all implemented using the D3.js~\citep{Bostock2011} library, while the statistical variables were generated in the backend by using Python. For the \textbf{Supplementary File} with additional analyses, please visit the paper's page\footnote{https://wilsonjr.github.io/cExpression}.

\section{Case studies}
\label{sec:case-studies}

To validate the proposed technique, we explore two document collections, a dataset of news articles from 2011 collected from different sources and a dataset of \textit{tweets} about COVID-19 symptoms collected inside the São Paulo state (Brazil) territory from March 2020 to August 2020. We also analyze multivariate data using a medical dataset in the \textbf{Supplementary File}.

\subsection{The news dataset}

\begin{figure}[!htb]
\centering
\includegraphics[width=0.4\linewidth]{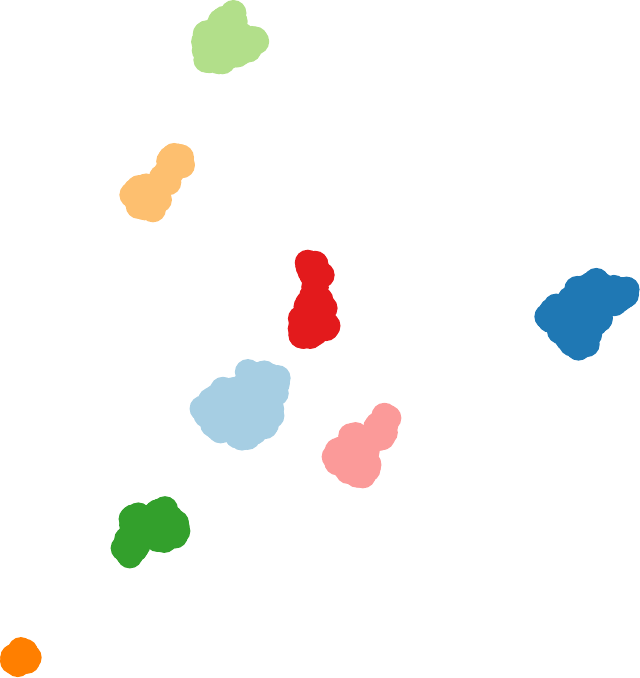}
\caption{UMAP projection of the \textit{news} dataset.}
\label{fig:news-overview}
\end{figure}

In this first case study, we inspect a document collection of $495$ news articles in English available in RSS format by Reuters, BBC, CNN, and Associated Press agencies. Fig.~\ref{fig:news-overview} shows the UMAP~\citep{McInnes2018} projection of the dataset color-coded based on the Leiden~\citep{Traag2018} algorithm. We used the first $40$ Principal Components (PCs) of the dataset to compute the neighborhood graph ($k = 15$) for the UMAP technique. The resolution parameter of the Leiden algorithm was set to $0.3$.

\begin{figure}[!htb]
\centering
\includegraphics[width=\linewidth]{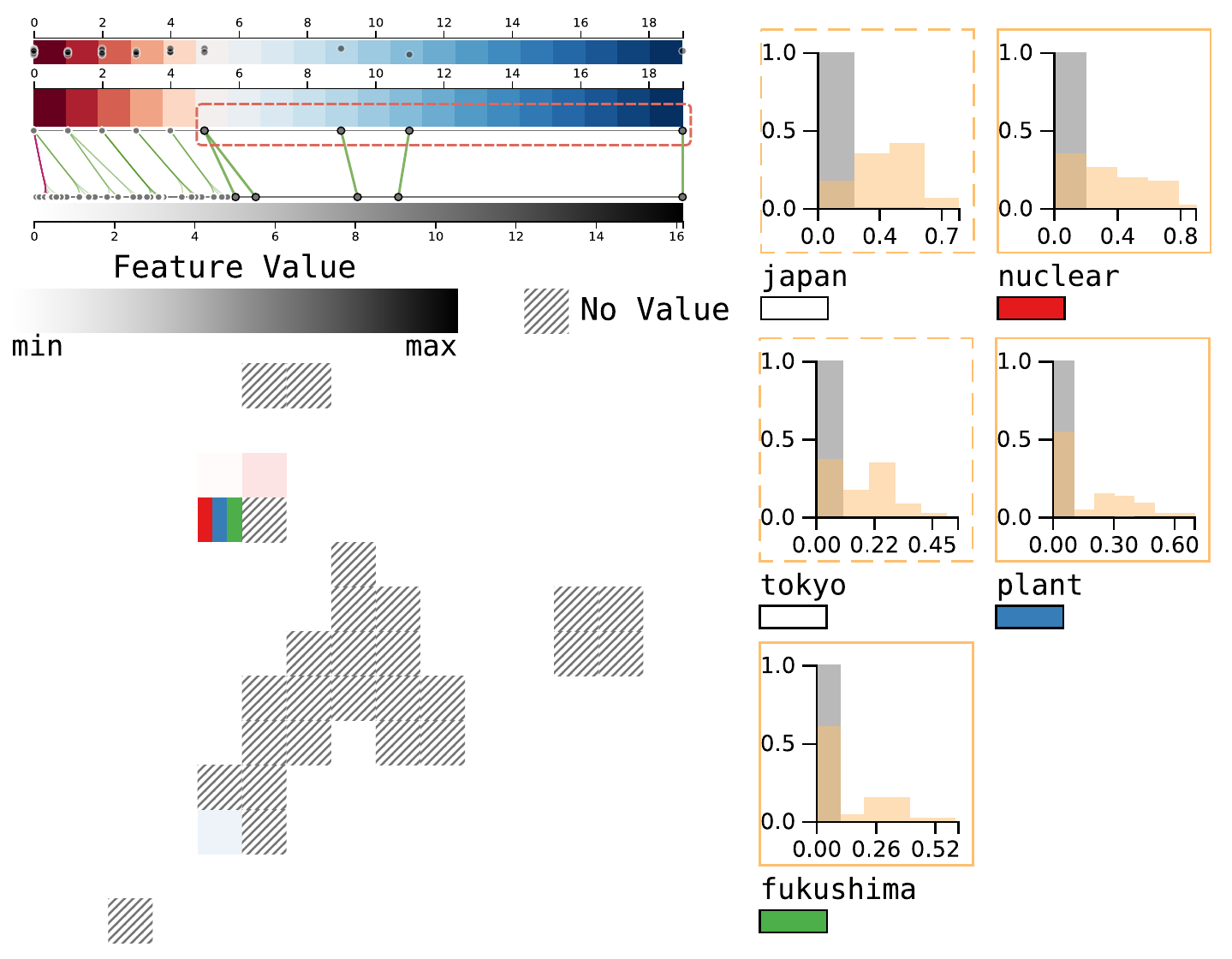}
\caption{After selecting the five most discriminative features, highlighting the terms \texttt{nuclear}, \texttt{plant}, and \texttt{fukushima} shows that UMAP distinguished the news articles separated the news articles related to the nuclear accident on the left side of the projection.}
\label{fig:news-light-orange1}
\end{figure}

Looking at the most distinctive terms presented in Fig.~\ref{fig:news-light-orange1}, and recalling that the news dataset contains articles from 2011, cluster \img{figs/case-studies/news/light-orange} represents news articles of the earthquake that hit Japan in 2011. The terms show that the incident on the nuclear plant of Fukushima I is present in almost all of the news articles due to the earthquake. Further, the news articles related to the nuclear power plant incident are concentrated in just one part of the cluster. The projection technique successfully uncovered a subcluster of news articles referring to the same aspects of the earthquake. 

\begin{figure}[!htb]
\centering
\includegraphics[width=\linewidth]{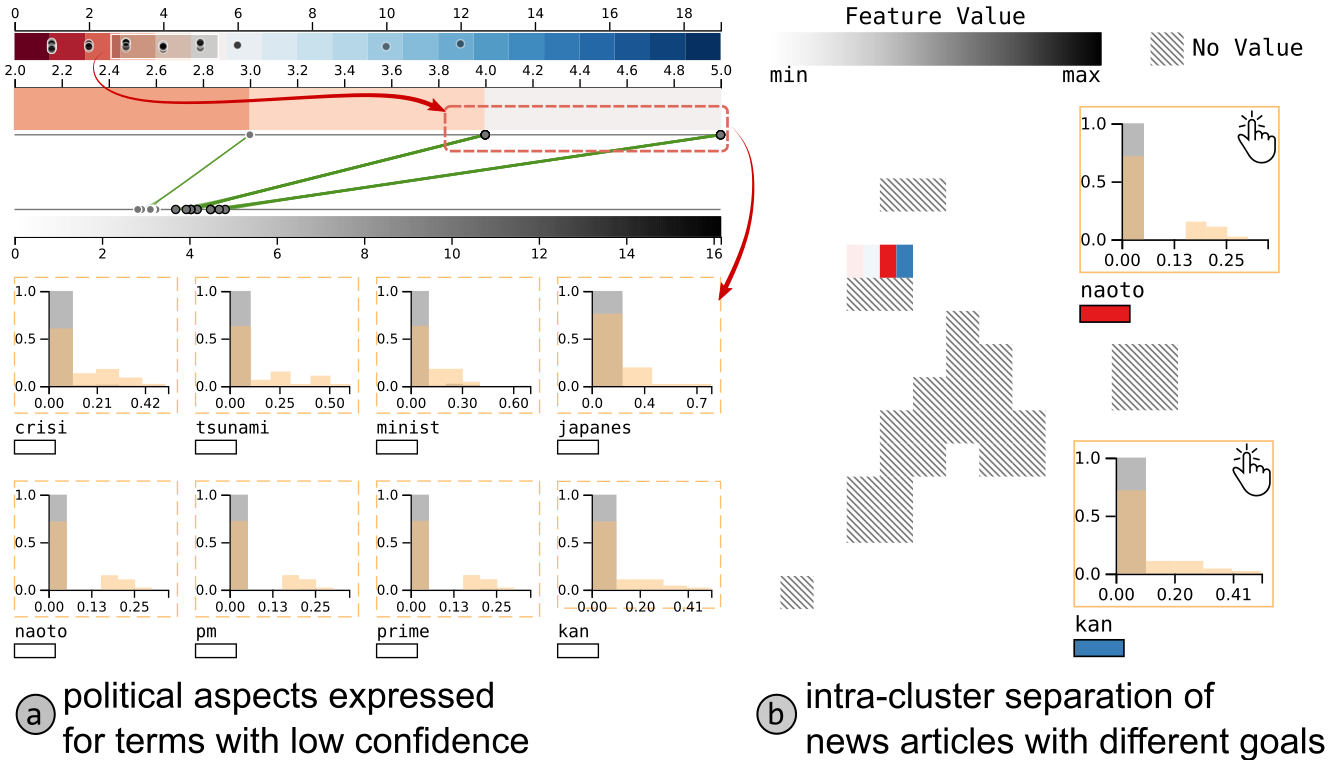}
\caption{Inspecting terms with lower confidence. Terms with lower confidence in the light-orange cluster are related to political aspects (a) and were positioned in a different subcluster.}
\label{fig:news-light-orange2-3}
\end{figure}

To further explore cluster \img{figs/case-studies/news/light-orange}, we focus on the terms that were not assigned much confidence (with $p$-values between $1\times10^{-3}$ and $1\times10^{-5}$) and then select a few terms, as illustrated in Fig.~\ref{fig:news-light-orange2-3} (a). In this case, the terms refer to former Japan's primer-minister, Naoto Kan, who resigned his role after the crisis provoked by the earthquake and the consequent incident in Fukushima I. The terms \texttt{naoto}, \texttt{kan}, \texttt{prime}, \texttt{minist}, \texttt{pm}, and \texttt{japanes} further support this idea.

Fig.~\ref{fig:news-light-orange2-3} (b) shows the scatterplot representation of the expression intensity of the terms \texttt{naoto} and \texttt{kan}, which validates our approach to understand the organization imposed by the dimensionality reduction technique. While news articles regarding the Fukushima I incident were positioned on the bottom of the cluster, news articles about the former primer-minister were positioned on the top of the cluster.

\begin{figure}[!htb]
\centering
\includegraphics[width=\linewidth]{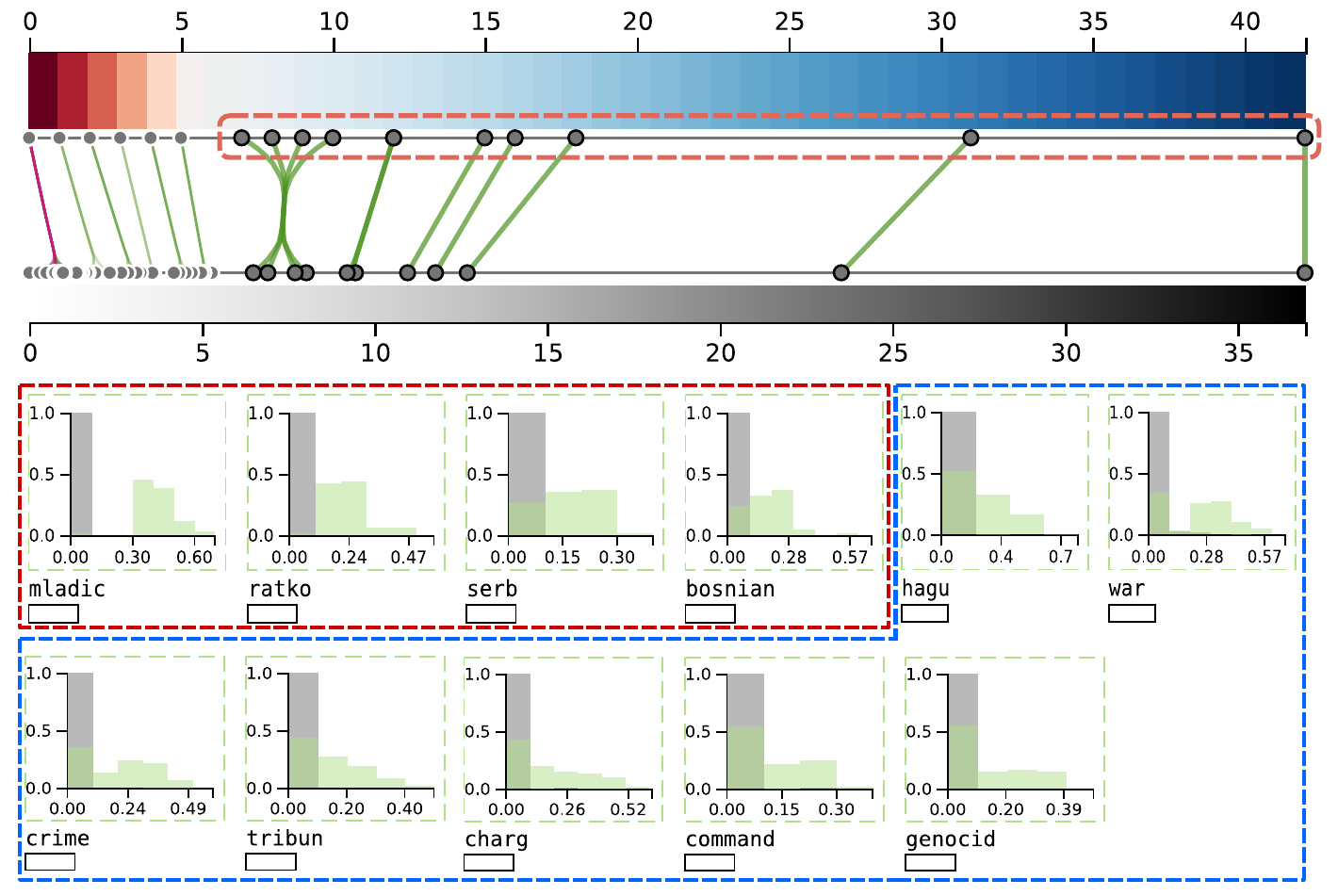}
\caption{Contrastive information showing levels of understanding. The terms highlighted in red show high-level information, while the terms highlighted in blue give specific hints about the news articles.}
\label{fig:news-light-green}
\end{figure}

Proceeding to cluster \img{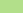} in Fig.~\ref{fig:news-light-green}, there are many terms with high confidence that could be used to understand the main topics of the news articles. The first terms (highlighted in red) correspond to Ratko Mladi\'{c}, a former Serbian military officer, head of the Serbian Republic Army during the Bosnian War between 1992-1995. The other group represents more specific information about the news articles contained in the cluster. The terms refer to the prison of Mladi\'{c} due to war crimes. More specifically, the terms \texttt{tribun} and \texttt{hagu} refer to the fact that Mladi\'{c} was extradited to The Hague city in the Netherlands to respond to his crimes in the International Court of Justice.

\begin{figure}[!htb]
\centering
\includegraphics[width=\linewidth]{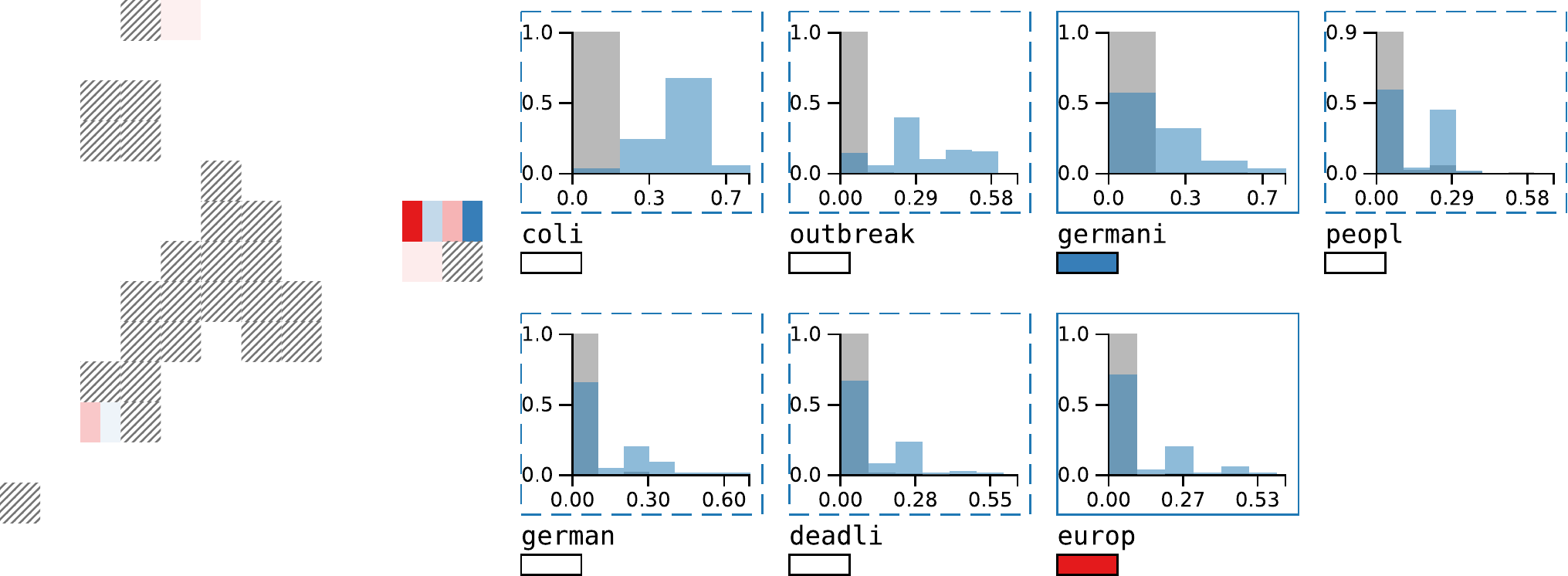}
\caption{Inspection of \texttt{germani} and \texttt{europ}'s terms shows that the dimensionality reduction separated news articles focused on the European continent from the news articles focused on Germany.}
\label{fig:news-blue}
\end{figure}

We finish the analysis by inspecting cluster \img{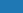}, as illustrated in Fig.~\ref{fig:news-blue}. From the terms highlighted as discriminative, such a cluster corresponds to news articles of a strain of Escherichia Coli O104:h4 bacteria outbreak in northern Germany from May to June 2011. The majority of the news articles mention \texttt{coli} and \texttt{outbreak}. While the terms \texttt{europ} and \texttt{germani} express on opposite sides of the cluster could indicate that some news articles focus specifically on Germany, and others referring to the whole of Europe. The other terms describe the whole cluster: \texttt{health}, \texttt{deadli}, and \texttt{infect}.

\subsection{Tweets of COVID-19 symptoms}

For this particular use case, we aim to analyze a complex document collection of \textit{tweets} about COVID-19 common symptoms (fever, high fever, cough, dry cough, difficulty breathing, shortness of breath) retrieved from São Paulo state (Brazil) from March 2020 to August 2020. To create the dataset, we retrieved tweets mentioning one of the symptoms discussed above; then, we classified the \textit{tweets} according to their relevance.  We used the BERT~\citep{Devlin2018} language model to train a classifier---\textbf{Supplementary File} contains all of the model's performance---to select only relevant \textit{tweets}.

\begin{figure}[!htb]
\centering
\includegraphics[width=0.3\linewidth]{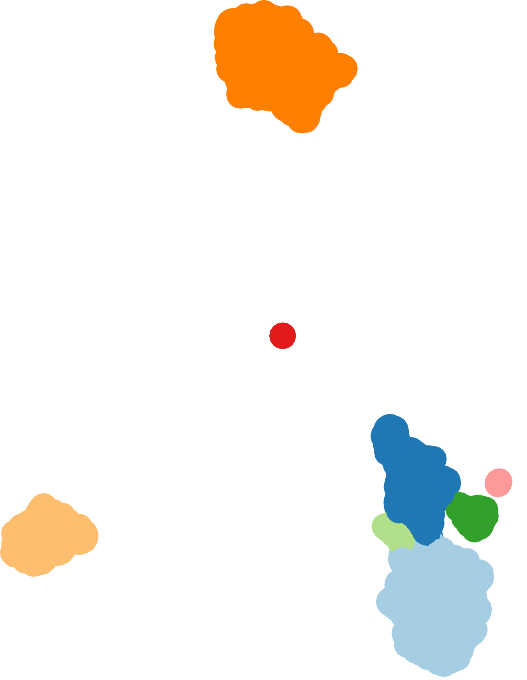}
\caption{UMAP projection of the \textit{tweets} about COVID-19.}
\label{fig:tweets-overview}
\end{figure}

We manually classified ten thousand \textit{tweets} as relevant or not and used the BERT model to automatically classify other 30 thousand \textit{tweets}, which we analyzed here. Relevant \textit{tweets} are those with serious comments about COVID-19 and where there is a chance of infection. Non-relevant \textit{tweets} correspond to news, jokes, and other informative comments. Further, to inspect the dataset projected on $\mathbb{R}^2$, we used the UMAP technique with the nearest neighbor graph set as 20. Finally, the clusters were manually defined, as shown in Fig.~\ref{fig:tweets-overview}.

The projection of the tweets shows four main clusters: red \img{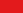}, dark orange \img{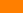}, orange \img{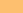}, and a cluster divided into five other subclusters (\img{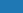}, \img{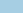}, \img{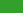}, \img{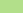}, \img{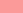}). To analyze the projection, we follow the strategy of analyzing the separated greater clusters and then proceeding to the very cohesive cluster \img{figs/case-studies/tweets/red}.

\begin{figure}[!htb]
\centering
\includegraphics[width=\linewidth]{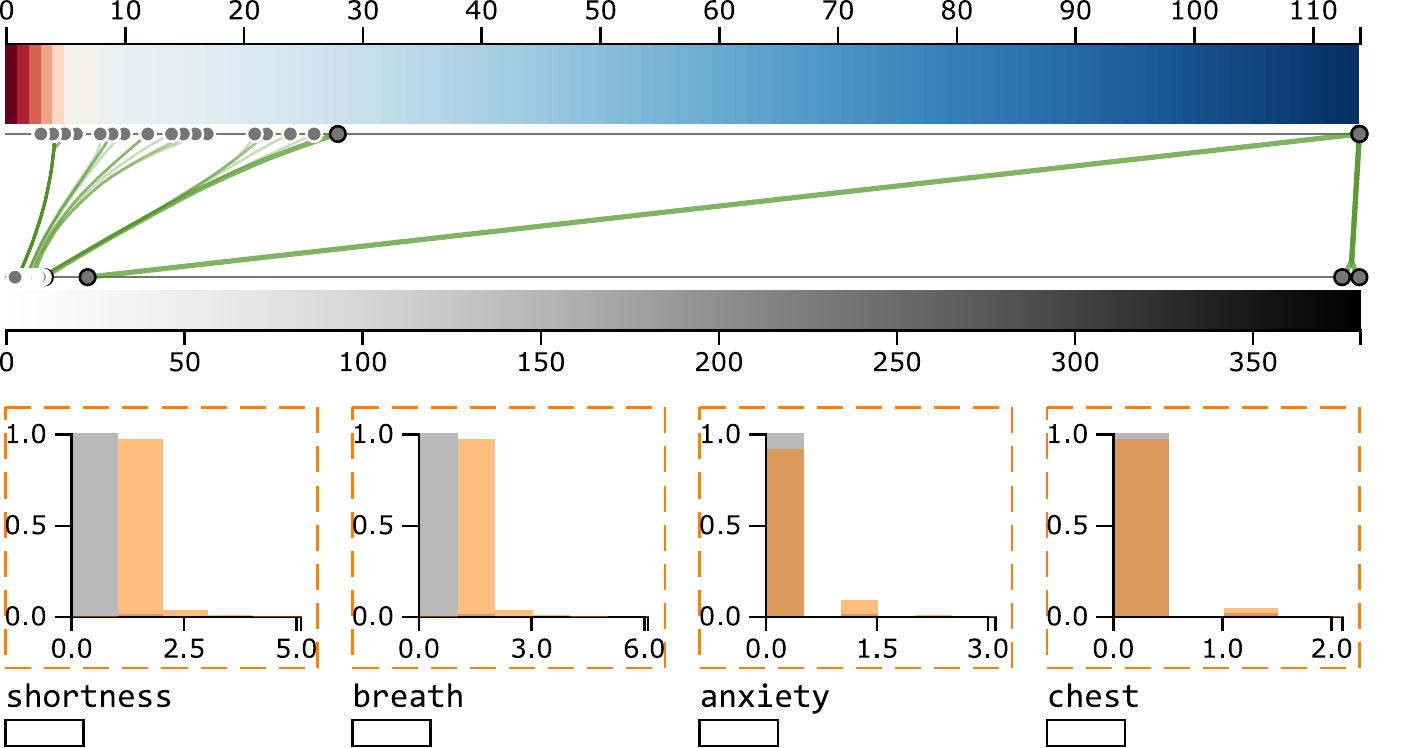}
\caption{Discriminative terms for the dark-orange cluster---terms related to respiratory problems.}
\label{fig:tweets-dark-orange1}
\end{figure}

Fig.~\ref{fig:tweets-dark-orange1} shows that cluster \img{figs/case-studies/tweets/dark-orange} is mainly related to tweets about the shortness of breath, one of the most severe COVID-19 symptoms---the term \texttt{chest}, for example, could indicate people describing how they were feeling and where the sensations are occurring in their body. Another interesting aspect of such a cluster is the term \texttt{anxiety}, which indicates a problem arising from the necessity of social isolation during the pandemic. In fact, during data reading, many tweets corresponded to people asking whether a COVID-19 infection or anxiety crisis caused their shortness of breath.

\begin{figure}[!htb]
\centering
\includegraphics[width=\linewidth]{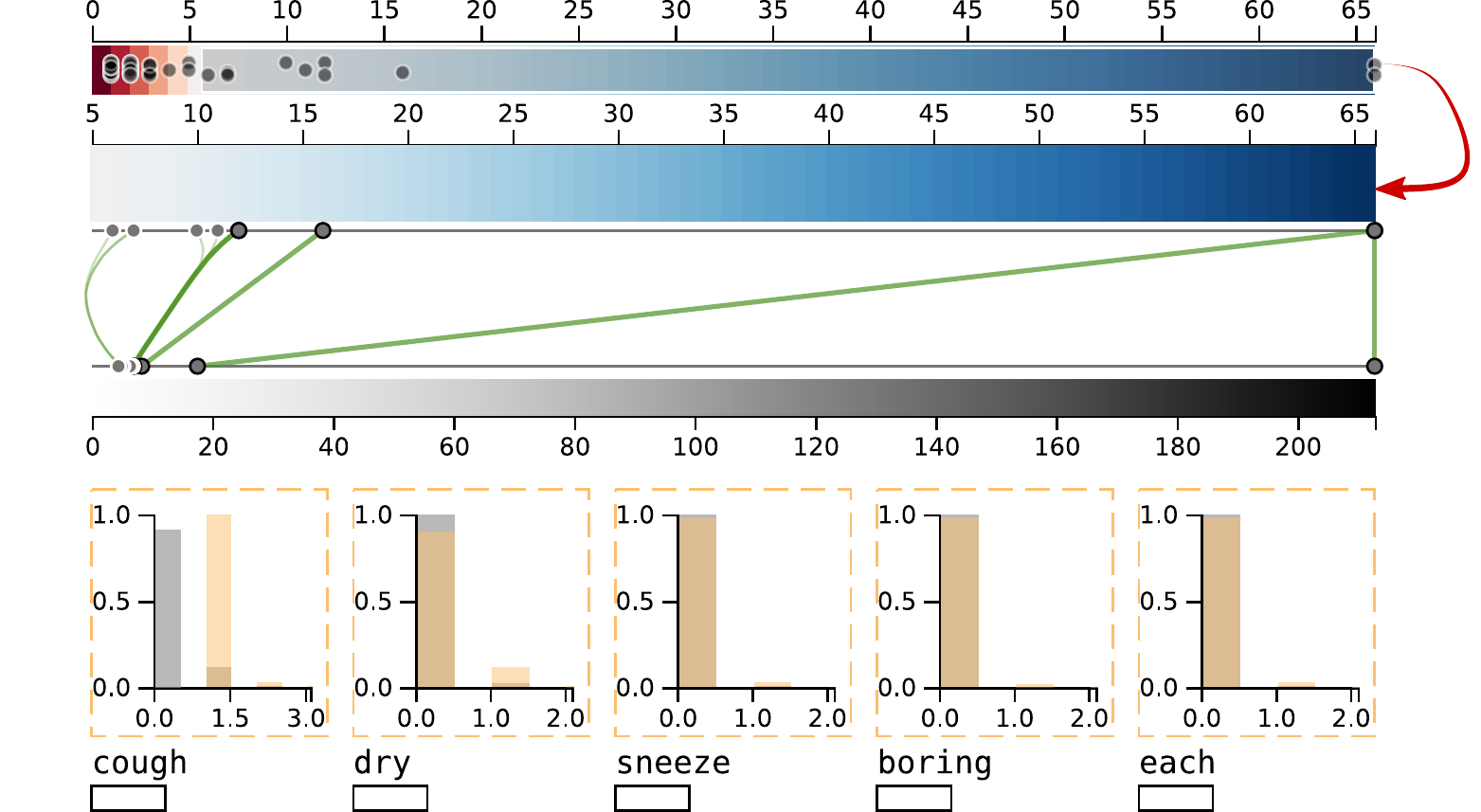}
\caption{Discriminative terms for the light-orange cluster. Terms indicate users tweeting their complaints about symptoms expressed during COVID-19 infection.}
\label{fig:tweets-orange1}
\end{figure}

Proceeding to cluster \img{figs/case-studies/tweets/orange}, we focus on the terms with p-value $\leq$ 1e-5. Fig.~\ref{fig:tweets-orange1} shows that such a cluster is related mainly to the \textit{dry cough} symptom of COVID-19 due to the presence of terms \texttt{cough} and \texttt{dry}---the term \texttt{sneeze} is also present when Twitter users were describing their symptoms. The other terms (\texttt{annoying} and \texttt{each}) usually consist of phrases commonly identified in the tweets, such as, (directly translated from informal Portuguese) ``I am with an annoying cough...'' or ``It is each cough...''.

\begin{figure}[!htb]
\centering
\includegraphics[width=\linewidth]{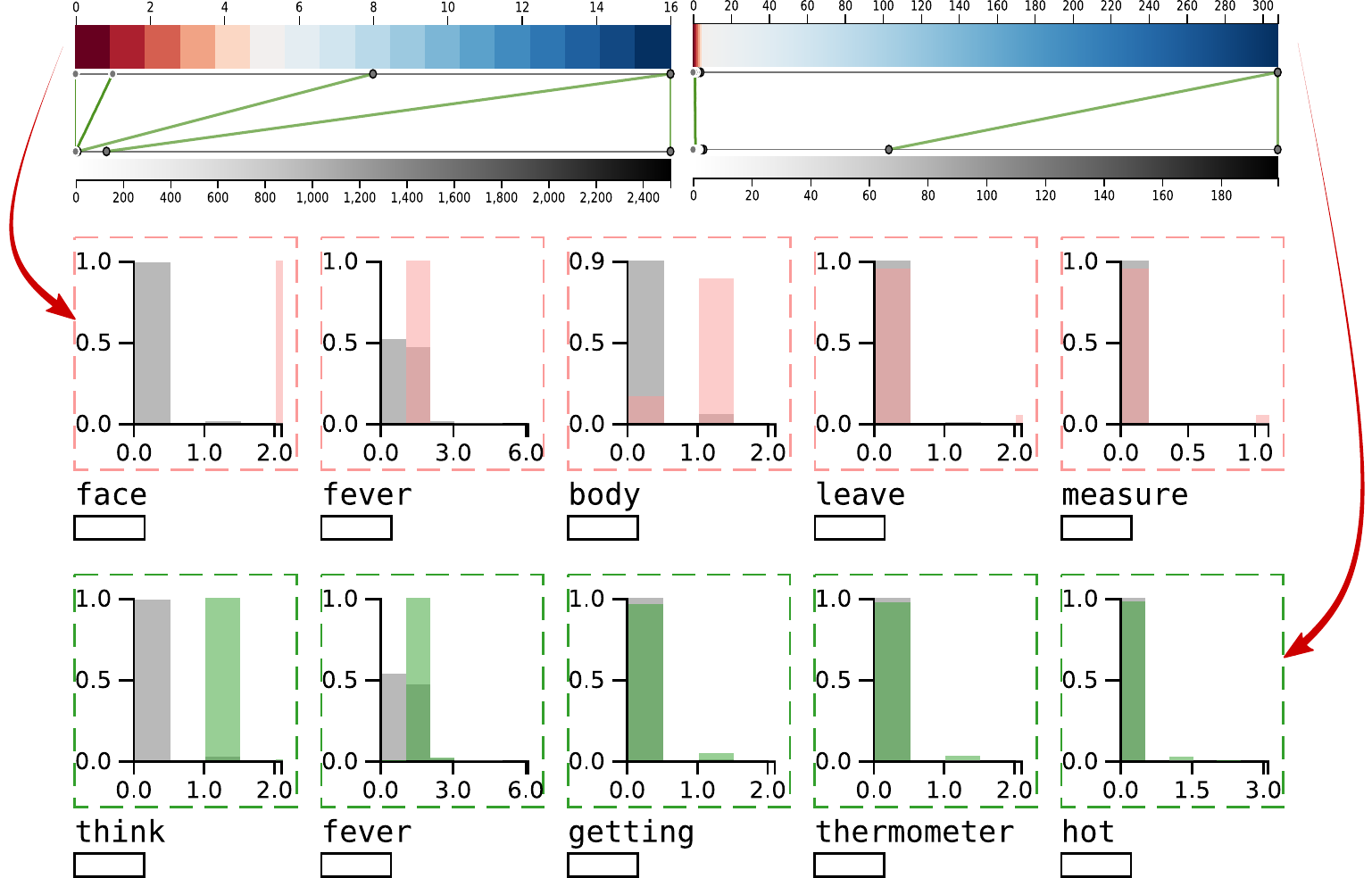}
\caption{Comparing two clusters with \texttt{fever} as the main subject.}
\label{fig:tweets-dark-green}
\end{figure}

Finally, we analyze the subclusters on the bottom right of Fig.~\ref{fig:tweets-overview}. As shown in Fig.~\ref{fig:tweets-dark-green}, both of clusters \img{figs/case-studies/tweets/dark-green} and \img{figs/case-studies/tweets/beige} refer to characteristics of fever symptom---see how the terms \texttt{face}, \texttt{fever}, \texttt{body}, and \texttt{to-measure} are expressed in cluster \img{figs/case-studies/tweets/beige} while the terms \texttt{I-think}, \texttt{fever}, \texttt{getting}, \texttt{thermometer},\ and \texttt{hot} are expressed in cluster \img{figs/case-studies/tweets/dark-green}. Particularly for cluster \img{figs/case-studies/tweets/dark-green}, the terms indicate users that had started to feel febrile when they posted the symptom on Twitter. Such insight could be significant to reveal which cities are presenting people with developing symptoms of COVID-19. That is, regulatory policies could be made based on the geolocalization of the \textit{tweets}.

\begin{figure}[!htb]
\centering
\includegraphics[width=\linewidth]{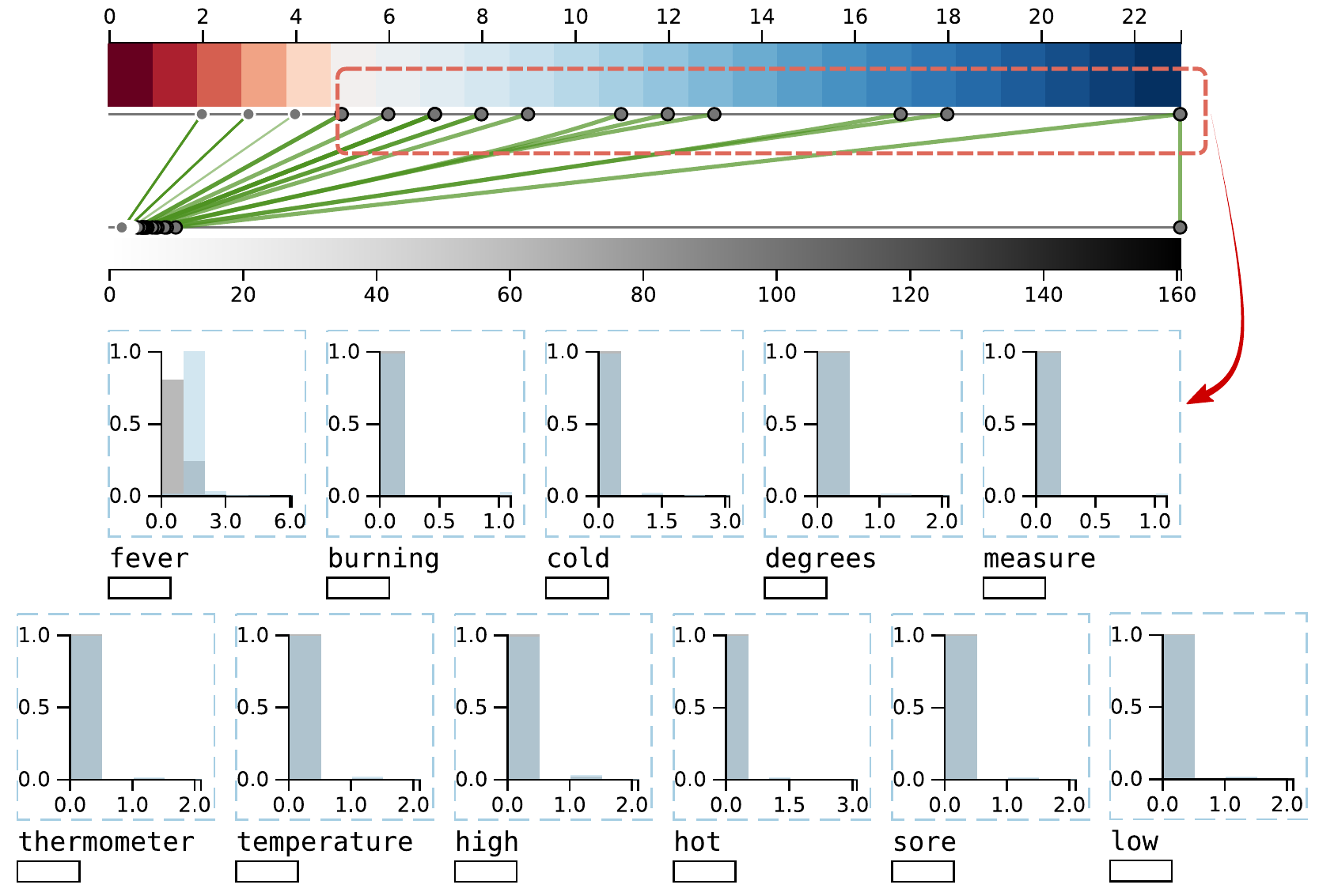}
\caption{Distinctive terms for the light-blue cluster. In this case, there are more general aspects of symptoms related to fever. For instance, the term \texttt{sore}, could be related to \textit{sore throat} or a new COVID-19 symptom.}
\label{fig:tweets-blue}
\end{figure}

Cluster \img{figs/case-studies/tweets/blue} indicates more general aspects about the symptoms related to fever. Fig.~\ref{fig:tweets-blue} shows that the discriminative terms are related to fever---notice that we omitted the terms \texttt{sick}, \texttt{was}, \texttt{dipyrona}, \texttt{pain}, \texttt{night}, \texttt{to-be}, and \texttt{bath} since they are terms used to create the phrases. An interesting aspect of this cluster is the presence of the term \texttt{sore}, an uncommon symptom.

\begin{figure}[!htb]
\centering
\includegraphics[width=\linewidth]{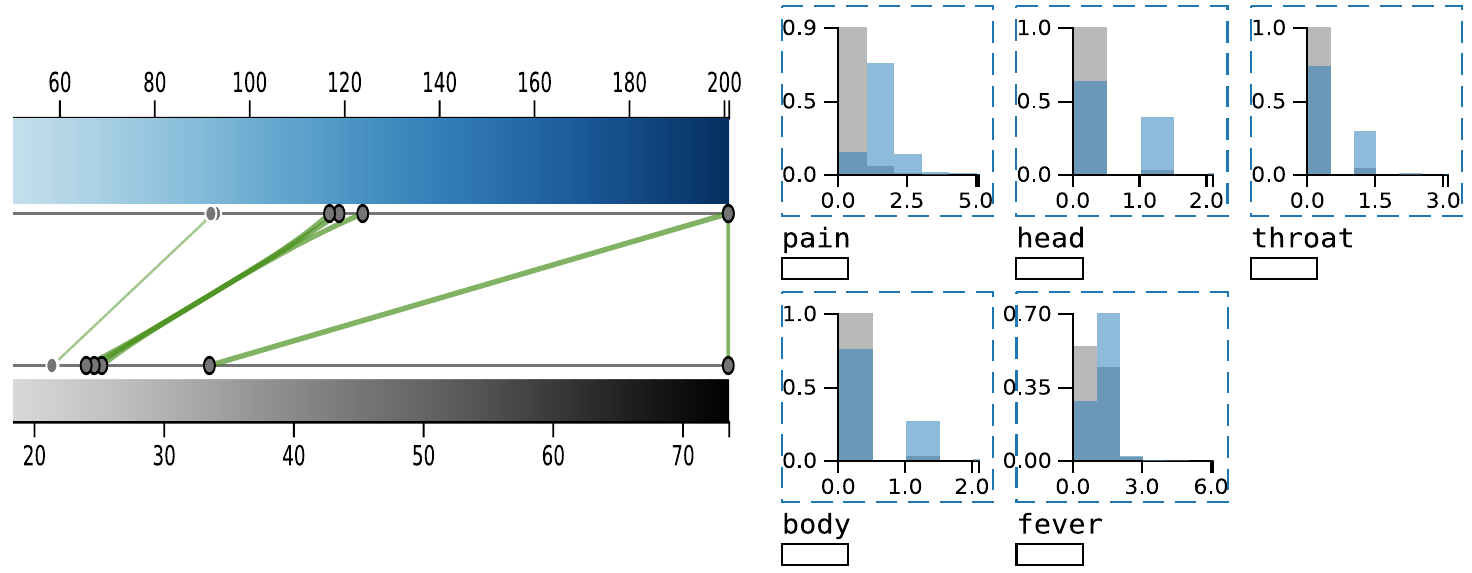}
\caption{Distinctive terms from the dark-blue cluster can be considered as indicative of Dengue symptoms.}
\label{fig:tweets-dark-blue1}
\end{figure}

Unlike the other clusters, cluster \img{figs/case-studies/tweets/dark-blue} presents many different levels of understanding. Firstly, Fig.~\ref{fig:tweets-dark-blue1} shows that although one of the five most discriminative refers to fever, such a cluster might represent Dengue symptoms. That is, while the citizens were posting how they were worried about COVID-19, the symptoms indicated on the tweets match with those used to describe Dengue infection (\texttt{pain}, \texttt{head}, \texttt{throat}, \texttt{body}, and \texttt{fever}), a common disease faced in Brazil.

\begin{figure}[!htb]
\centering
\includegraphics[width=\linewidth]{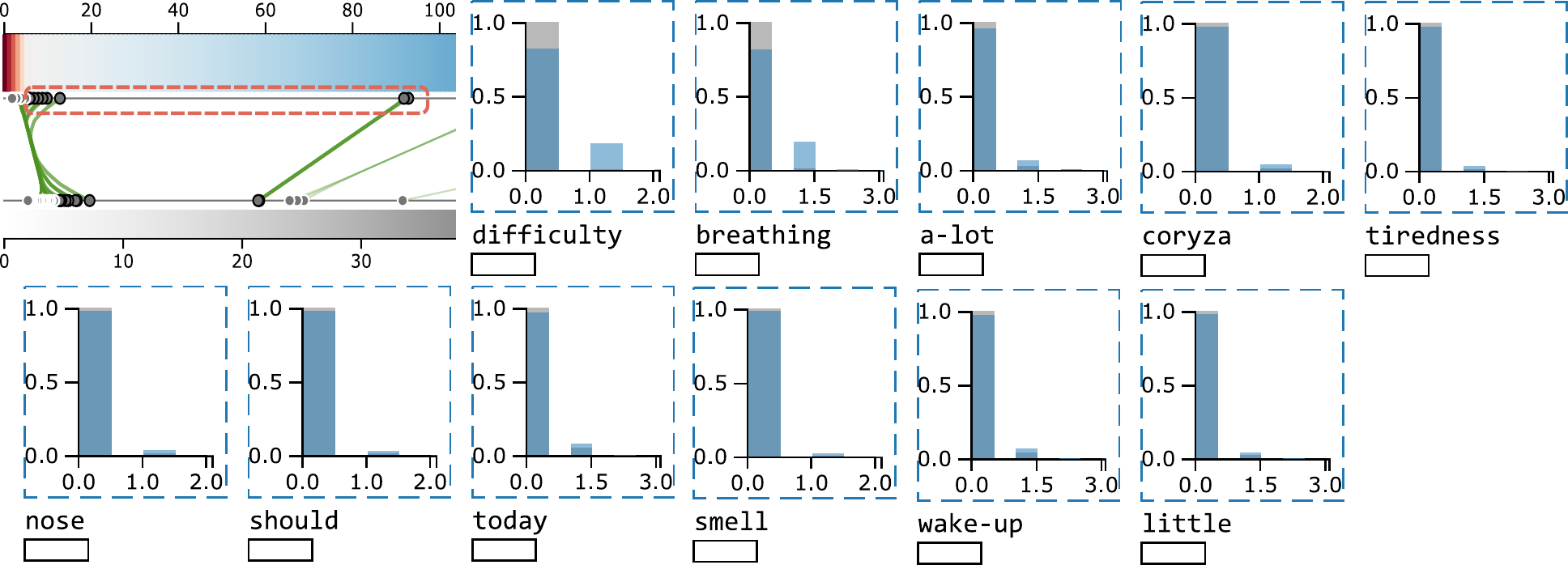}
\caption{Evaluating another term from the dark-blue cluster. The terms such as \texttt{difficulty} and \texttt{breathing} indicate the most severe COVID-19 symptoms, shortness of breath.}
\label{fig:tweets-dark-blue2}
\end{figure}

Finally, Fig.~\ref{fig:tweets-dark-blue2} shows that a severe symptom of COVID-19 is also present in cluster \img{figs/case-studies/tweets/dark-blue}: \texttt{difficulty breathing}. Besides that, other terms indicate common symptoms of seasonal flu, such as \texttt{coryza}. Finally, the term \texttt{tiredness} reinforces the idea of Dengue symptoms or consists of a uncommon COVID-19 symptom.

\begin{figure}[!htb]
\centering
\includegraphics[width=\linewidth]{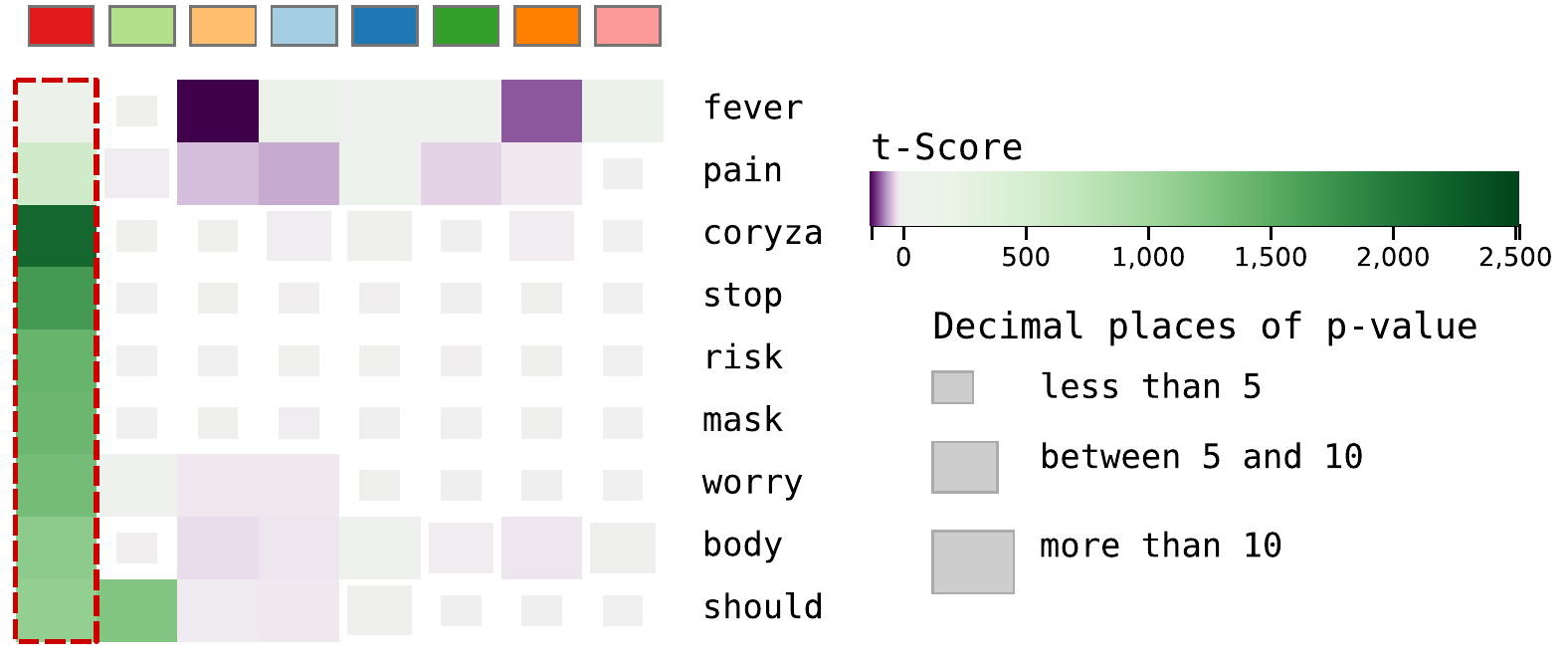}
\caption{Heatmap showing that the red cluster comprises the tweets where people are worried about wearing masks and coryza.}
\label{fig:tweets-red}
\end{figure}

In this case study, the tweets presented three main topics: clusters \img{figs/case-studies/tweets/orange} is related to respiratory problems induced by COVID-19 infection, as well as related to anxiety crisis due to long periods of social isolation; cluster \img{figs/case-studies/tweets/orange} is related to \textit{dry cough}, a common COVID-19 symptom; finally, the cluster with subclusters has a lot of aspects regarding high fever, where each one of the subcluster (\img{figs/case-studies/tweets/dark-blue}, \img{figs/case-studies/tweets/dark-green}, \img{figs/case-studies/tweets/blue}, \img{figs/case-studies/tweets/green}, \img{figs/case-studies/tweets/beige}) has its particularities. Finally, to investigate cluster \img{figs/case-studies/tweets/red}, the heatmap shown in Fig.~\ref{fig:tweets-red} shows much of the very cohesive cluster is comprised of tweets concerning terms of worrying (see \texttt{worry} and \texttt{risk}) about coryza and wearing masks.

\section{Evaluation}
\label{sec:evaluation}

To further testify \textit{cExpression} to support analysis of dimensionality reduction results, we compare it against well-known topic extraction techniques using the \textit{cohesion} metric. Finally, we assess run-time execution of \textit{cExpression} and \textit{ccPCA}~\citep{Fujiwara2019} upon various dimensionality values for a document collection.

\subsection{Cohesion}

\begin{figure}[!htb]
\centering
\includegraphics[width=\linewidth]{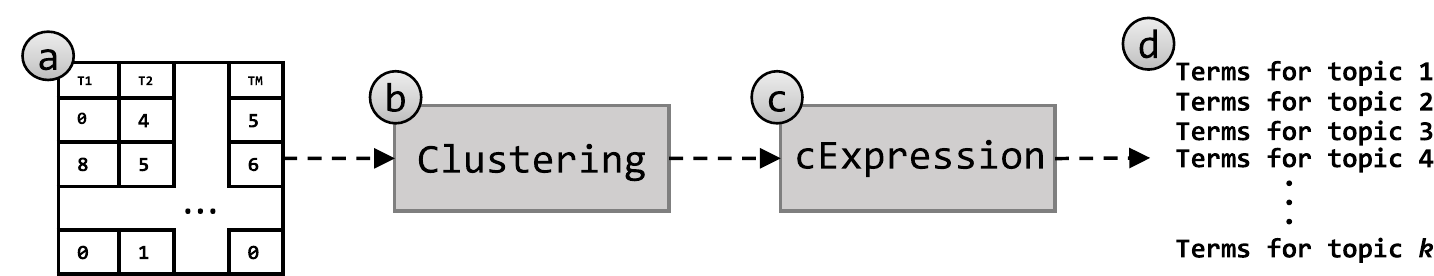}
\caption{Retrieving topics using cExpression. Given a bag-of-words representation of a document collection (\textbf{a}), we compute the dataset clusters (\textbf{b}). \textit{cExpression} is executed to retrieve distinctive terms for each cluster (\textbf{c}), and the topics correspond to the first $m$ terms of each cluster.}
\label{fig:diagram-topic}
\end{figure}

A useful way to analyze \textit{cExpression} results is by assessing how the terms selected from a document collection describe the clusters, 
similarly as in topic extraction tasks. Fig.~\ref{fig:diagram-topic} shows the pipeline for such a task.

After creating a bag-of-words representation (\textbf{a}.) of a document collection, we apply a clustering algorithm to group documents based on similarity (\textbf{b}.). Then, \textit{cExpression} returns the terms in each cluster that are not likely to appear in the other clusters. As explained in Section~\ref{sec:methodology-motivation}, these terms are ordered based on their contrastive score---the first $m$ terms are returned to compose the topics of each cluster (\textbf{d.}).

We use the \textit{Topic coherence}~\citep{Roder2015} metric to evaluate the robustness of the terms returned by \textit{cExpresion}. This metric is applied to the top $m$ terms of a topic and consists of the average of the pairwise word-similarity scores of the topic. A good model will generate coherent topics with high topic coherence scores. Thus, the topic extraction technique \textbf{A} is better than the technique \textbf{B} if it has a greater score.

The evaluation was performed using three datasets: $495$ news articles ($205$ terms) and $40794$ tweets related to COVID-19 symptoms ($295$ terms), and the \textit{20newsgroups}\footnote{http://qwone.com/~jason/20Newsgroups/} dataset. For the \textit{20newsgroups}, we use a subset comprising the classes \texttt{alt.atheism}, \texttt{comp.graphics}, \texttt{comp.windows.x}, \texttt{rec.motorcycles}, \texttt{sci.electronics}, \texttt{sci.med}, \texttt{talk.politics.guns}, \texttt{talk.politics.misc}, \texttt{talk.religion.misc}. After preprocessing the dataset---removal of English stop-words and terms appearing in a proportion of documents below 0.5---the dataset resulted in a bag-of-words representation of $4828$ documents by $3347$ terms. Finally, we compared \textit{cExpression} against the well-known and commonly used topic extraction approaches, LDA~\citep{Hoffman2010} and NMF~\citep{Cichocki2009}.

\begin{figure}[!htb]
\centering
\includegraphics[width=\linewidth]{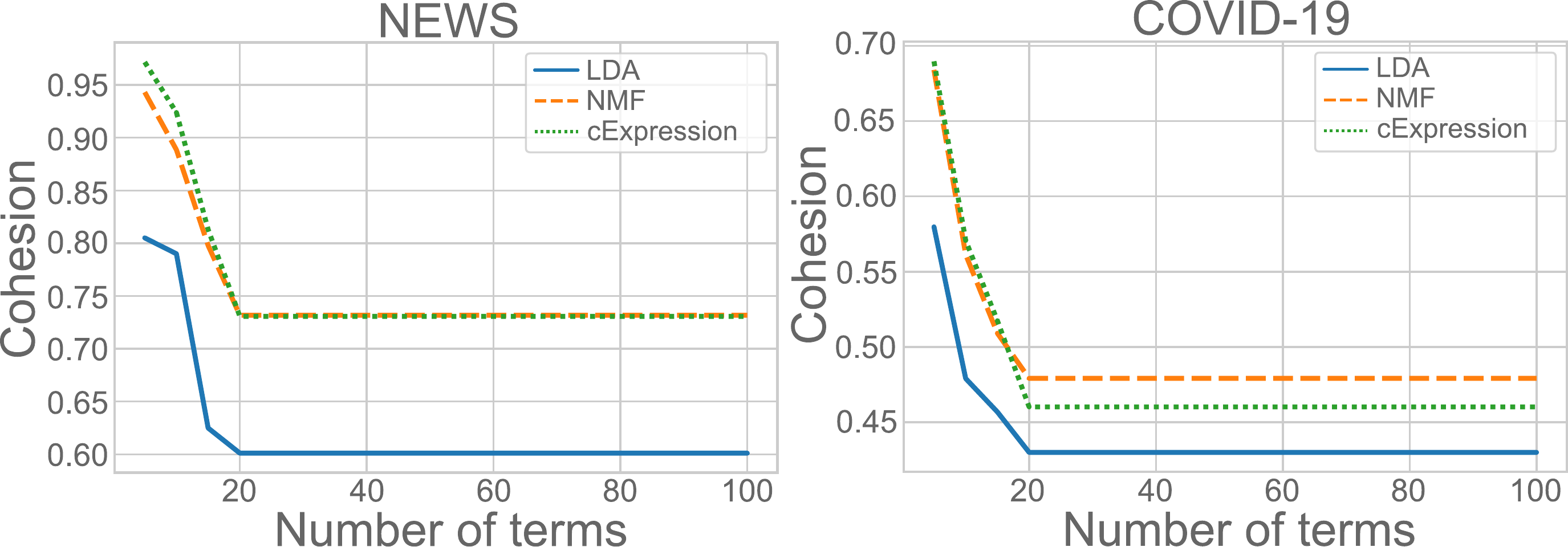}
\caption{cExpression shows competitive results for topics with a low number of terms ($\leq 20$).}
\label{fig:coherence_news_twitter}
\end{figure}

Fig.~\ref{fig:coherence_news_twitter} shows the cohesion when varying the number of terms in each topic for the \textit{news} and \textit{covid-19} datasets. \textit{cExpression} presents a slight advantage for topics with size below 20 and emphasizes the terms that are not expressed in other clusters---terms of topic A (cluster A) are not likely to appear in other topics. So, for a lower number of terms, \textit{cExpression} emphasizes the differences while showing good coherence.

To analyze the \texttt{20newsgroups} dataset, besides varying the number of terms, we also varied the number of topics returned from the algorithm. As shown in Table~\ref{tab:auc_20newsgroups}, we computed the Area Under the Curve (AUC) to summarize the results. \textit{cExpression} surpasses LDA and NMF for the most number of topics---it only presents lower AUC for two and four topics.

\begin{table}[!htb]
\centering
\begin{tabular}{c|c|c|c}
\hline
Number of topics & LDA     & NMF     & cExpression \\ \hline
2                & 41.0876 & 21.3630 & 32.9847         \\ \hline
3                & 38.3072 & 28.9405 & 39.0384         \\ \hline
4                & 37.7737 & 28.8734 & 37.4541         \\ \hline
5                & 37.9169 & 28.9459 & 38.2428         \\ \hline
6                & 37.5037 & 27.7676 & 41.6532         \\ \hline
7                & 37.4346 & 26.7356 & 40.2814         \\ \hline
8                & 33.4367 & 28.1161 & 39.3023         \\ \hline
9                & 33.2409 & 26.7829 & 36.5034         \\ \hline
10               & 35.5672 & 26.1085 & 37.2073         \\ \hline
\end{tabular}
\caption{Summarization of results in Cohesion using AUC. cExpression only returns lower results for two and four topics.}
\label{tab:auc_20newsgroups}
\end{table}

\begin{figure}[!htb]
\centering
\includegraphics[width=\linewidth]{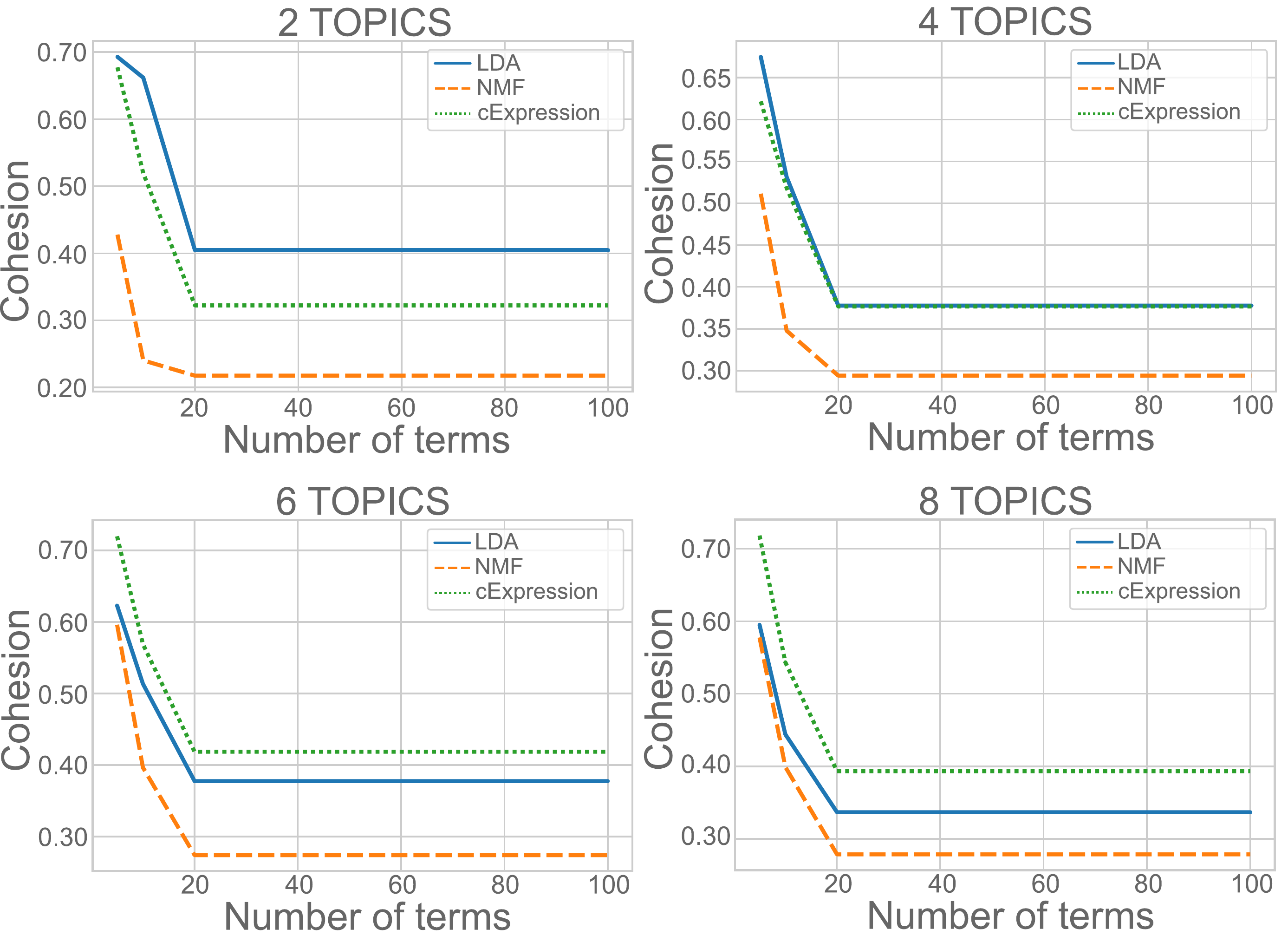}
\caption{cExpression was able to surpass LDA and NMF for the majority of the number of topics.}
\label{fig:coherence_20newsgroups}
\end{figure}

Fig.~\ref{fig:coherence_20newsgroups} shows that for only two topics, our technique could not uncover topics as good as LDA. Using four topics, our technique presents slightly lower results when using a few terms (below ten terms). Again, \textit{cExpression} performs better for only a few terms in the topic (six and eight). The remaining of the plots are in the \textbf{Supplementary File}.

\subsection{Run-time execution}

\begin{figure}[h!]
\centering
\includegraphics[width=\linewidth]{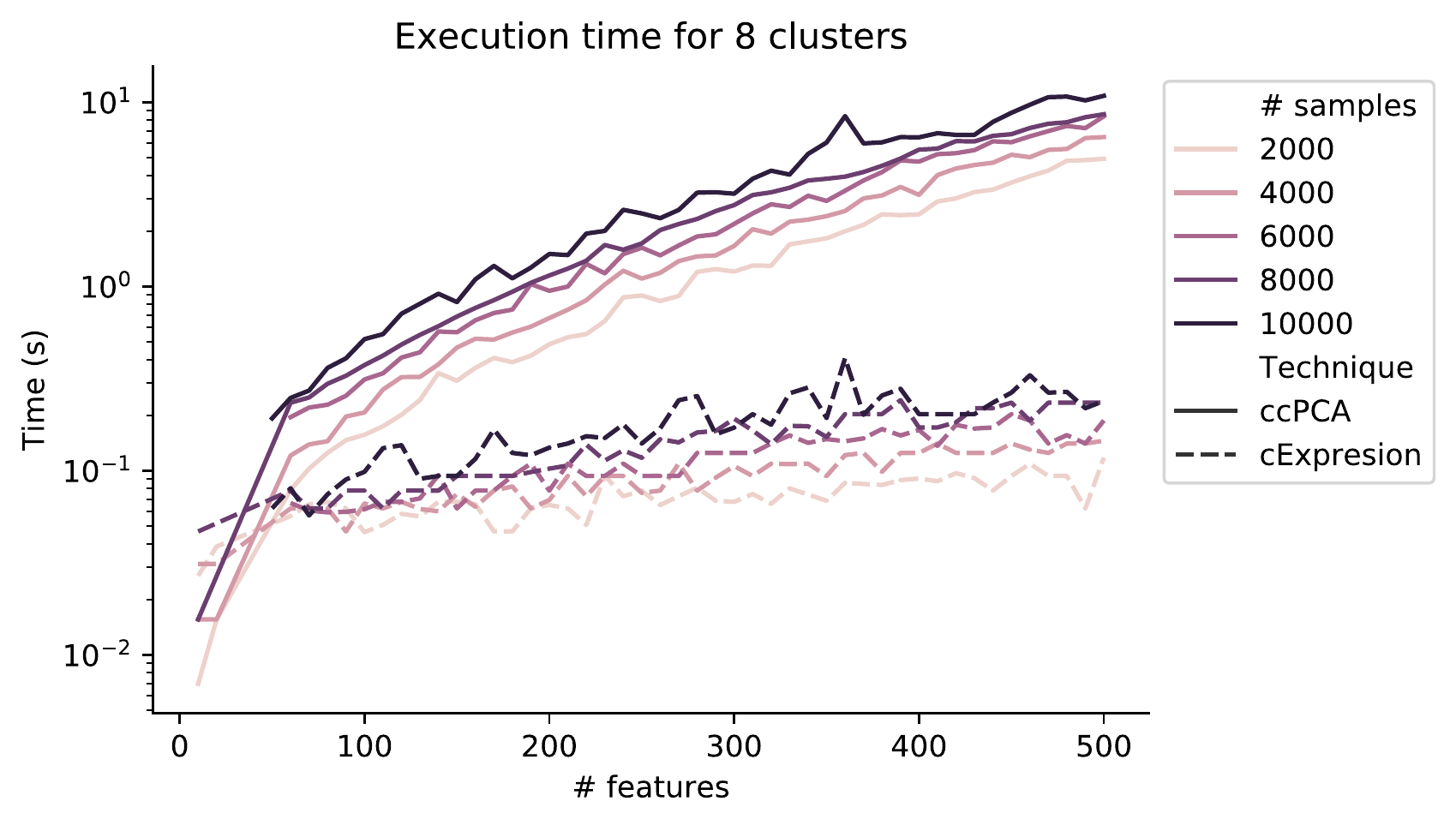}
\caption{Time execution in logarithmic scale for retrieving distinctive features using contrastive analysis.}\label{fig:time_execution}
\end{figure}

A technique must cope with high dimensionality to successfully analyze real-world data. Thus, we analyze the run-time execution of \textit{cExpression} and \textit{ccPCA} by varying the number of features and samples to perform analysis---we focus only on the techniques used to retrieve contrastive information. The experiment was performed in a dataset composed of 40794 tweets about COVID-19 symptoms (see Section~\ref{sec:case-studies} for a detailed description of a smaller version of it). Then, to evaluate the techniques, we generated different versions of the dataset by ranging the number of features between ten and 2000 and the number of samples between 2000 and 40000. Fig.~\ref{fig:time_execution} shows that, for eight clusters, \textit{cExpression} is faster than \textit{ccPCA} when we augment the dataset dimensionality. The \textbf{Supplementary File (Fig. 5)} also shows run-time executions for different number of clusters presenting the same pattern of Fig.~\ref{fig:time_execution}, which testify the superiority of \textit{cExpression} over \textit{ccPCA} regarding run-time execution.


\section{Discussions and Limitations}
\label{sec:discussions}

Contrastive analysis of dimensionality reduction results offers important mechanisms to understand how clusters differ in the projected space. Although the literature already presents a method for this task, we demonstrated that interactive visualizations with well-known statistics can enhance interpretation of the differences among clusters. There are a few other aspects regarding our approach, which we discuss in the following. 


\textbf{Cell-based encoding of scatterplot}. To facilitate local and global analyses among data points, the distribution of feature values visualized on the scatterplot helps analysts gain insight into the datasets structures. Such characteristic also reduces change blindness since uses are not required to change between color scales to visualize the feature value distribution on the scatterplot.

\textbf{Usability}. In the use cases, we evaluate the proposed tool regarding its strengths to help understand the cluster formation after dimensionality reduction. However, we plan to evaluate the tool's usability through user experiments to understand how it benefits analysis in well-defined tasks.

\textbf{Analyzing different DR techniques}. An important aspect of cExpression is how its results vary for different DR reduction techniques. In \textbf{Supplementary File}, we provide two additional analyzes on this characteristic. The results show, as expected, that if the clusters are the same---i.e., same data points in the same clusters---, the organization of the projected space does not matter, which can be explained by the fact that we feed the same data subsets to cExpression. We plan to use $k$ nearest neighbors to define the subsets, making the results more dependent on the projection structures than projection clusters.

\textbf{Features ordering}. Particularly for datasets where numbers indicate the presence of elements, such as in document collections where the bag-of-words representation indicates a term in a document, we \textit{do not} order the features based on the absolute value of their \textit{t}-score. By not ordering using absolute value, the features on the first positions (notice that we order in decreasing way) consist of the features present on clusters of interest. That is, features with negative \textit{t}-scores for such scenarios are the ones not in the cluster.

Although our work addresses important problems in the literature, there are few other aspects that need further research.

\textbf{Visual scalability of the number of clusters}. In Section~\ref{sec:feature-visualization}, we comment on the visual scalability related to the dataset dimensionality. Another point of view is related to the number of clusters, in which the higher the number of clusters more iterations the user would have to perform to cover all of the dataset. While we do not address this problem in this work, our methodology is fast enough not to add a burden during analysis. Further that, we plan to investigate how multilevel dimensionality reduction strategies~\cite{Pezzotti_2016, MarcilioJr2021_ExplorerTree, MarcilioJr2021_humap} would help to alleviate the analysis by applying overview first and details on-demand operations.

\textbf{Visual scalability of the number of features}. When visualizing the distribution of a greater number of features on the scatterplot view, by augmenting the boxes' area to circumvent the space dedicated to visualizing feature intensity, users can better distinguish among different classes. However, such an increase can remove the scatter-plot context. The number of simultaneous features visualized can also impact the ability of users to distinguish the different intensities among the features. We plan to investigate glyph-based visualizations to encode more information---for example, using star plot and color encoding to visualize feature values and colors, respectively.

\textbf{Assumptions for the t-Student test}. Although t-test provides the dataset features' discriminative power, we have assumed the samples follow normal distributions. Besides that, computing the \textit{t}-statistic can be problematic because the variance estimates can be skewed by features having a very low variance~\citep{Jeanmougi2010}. Consequently, these features are associated with a large t-statistic and falsely selected discriminative terms~\citep{Tusher2001}. Another drawback comes from its application on small sample sizes, which implies low statistical power~\citep{Murie2008}. One way to address these limitations is already in our tool. It allows users to investigate the distribution of values in the scatterplot representation using the color scale or to use the encoding to visualize various features simultaneously.

\textbf{Selection of clusters in the visual space}. As discussed in the paper, we focus on explaining the visual clusters---i.e., the clusters perceived in the visual space. Thus, the input data containing the cluster labels must match the clusters in the visual space for trustworthy and consistent exploratory analysis. When this is not true, and cluster labels are assigned to different clusters in the visual space, the cExpression technique will not identify truly distinctive features since the feature value distribution would be similar (see Section~\ref{sec:methodology-motivation}). However, this limitation is easily overcome by preprocessing the input file with a consistent assignment of cluster labels.

\section{Conclusion}
\label{sec:conclusion}

Understanding the influence of features on the formation of clusters and sub-clusters is a promising approach when analyzing dimensionality reduction results represented by scatterplots. Existing methods to address this task emphasize global characteristics not capable of differentiating clusters. On the other hand, current methods for contrastive analysis need unrealistic run-time execution for practical applications.

This paper presents a novel approach for contrastive analysis of dimensionality reduction results represented by scatterplots. We use a bipartite graph metaphor to represent the relation among statistics (t-score and p-value) of each cluster feature. Using focus+context analysis, we show how our approach can retrieve insights about cluster organization even for complex datasets. Finally, we also show how such an approach can be robust by comparing it against well-known topic extraction techniques.

\section*{Acknowledgements}

This research work was supported by FAPESP (São Paulo Research Foundation), grants \#2018/17881-3 and \#2018/25755-8, and by the Coord. de Aperfeiçoamento de Pessoal de Nível Superior - Brazil (CAPES), grant \#88887.487331/2020-00. We also thank the anonymous reviewers for their valuable comments.

\bibliographystyle{elsarticle-num}

\bibliography{cas-refs}


\end{document}